\documentclass{JHEP3}
\usepackage{units}
\usepackage{graphicx,epsfig}
\usepackage{amsmath}

\newcommand{\lsim}{\lesssim}
\newcommand{\gsim}{\gtrsim}

\newcommand{\Ai}{\mathop\mathrm{Ai}}
\newcommand{\Bi}{\mathop\mathrm{Bi}}
\newcommand{\Tr}{\mathop\mathrm{Tr}}

\newcommand{\sign}{\mathop\mathrm{sign}}
\newcommand{\vev}[1]{\langle #1 \rangle}

\newcommand{\bk}{\mathbf{k}}
\newcommand{\be}{\begin{equation}}
\newcommand{\ee}{\end{equation}}
\newcommand{\ba}{\begin{eqnarray}}
\newcommand{\ea}{\end{eqnarray}}
\let\rmi=\mathrm

\renewcommand{\l}{\left(}
\renewcommand{\r}{\right)}

\newcommand{\half}{\frac{1}{2}}

\newcommand{\cL}{\mathcal{L}}

\newcommand{\e}{\mathrm{e}}
\newcommand{\const}{\mathop\mathrm{const}}
\newcommand{\nuMSM}{\ifmmode\nu\mathrm{MSM}\else$\nu$MSM\fi}
\newcommand{\dm}{\partial_\mu}

\newcommand{\Rk}{R_\bk}

\newcommand{\tcr}{t_{\mathrm{cr}}}
\newcommand{\Xcr}{X_{\mathrm{cr}}}

\title{On initial conditions for the Hot Big Bang}

\author{F. Bezrukov\\
  Max-Planck-Institut f\"ur Kernphysik,\\
  PO Box 103980, 69029 Heidelberg, Germany\\
  Institute for Nuclear Research of the Russian Academy of Sciences,\\
  60th October Anniversary prospect 7a, Moscow 117312, Russia\\
  E-mail: \email{Fedor.Bezrukov@mpi-hd.mpg.de}}
\author{D. Gorbunov\\
  Institute for Nuclear Research of the Russian Academy of Sciences,\\
  60th October Anniversary prospect 7a, Moscow 117312, Russia\\
  E-mail: \email{gorby@ms2.inr.ac.ru}
}
\author{M. Shaposhnikov\\
  Institut de Th\'eorie des Ph\'enom\`enes Physiques,\\
  \'Ecole Polytechnique F\'ed\'erale de Lausanne,\\
  CH-1015 Lausanne, Switzerland    \\
  E-mail: \email{Mikhail.Shaposhnikov@epfl.ch}
}

\abstract{We analyse the process of reheating the Universe in the
  electroweak theory where the Higgs field plays a role of the
  inflaton.  We estimate the maximal temperature of the Universe and
  fix the initial conditions for radiation-dominated phase of the
  Universe expansion in the framework of the Standard Model (SM) and of the
  $\nu$MSM~--- the minimal extension of the SM by three
  right-handed singlet fermions.  We show that the inflationary epoch
  is followed by a matter dominated stage related to the Higgs field
  oscillations.  We investigate the energy transfer from
  Higgs-inflaton to the SM particles and show that the radiation
  dominated phase of the Universe expansion starts at temperature $T_r
  \simeq (3-15) \times \unit[10^{13}]{GeV}$, where the upper bound
  depends on the Higgs boson mass.  We estimate the
  production rate of singlet fermions at preheating and find that
  their concentrations at $T_r$ are negligibly small.  This suggests
  that the sterile neutrino Dark Matter (DM) production and
  baryogenesis in the $\nu$MSM with Higgs-driven inflation are low
  energy phenomena, having nothing to do with inflation.  We study
  then a modification of the $\nu$MSM, adding to its Lagrangian higher
  dimensional operators suppressed by the Planck scale.  The role of
  these operators in Higgs-driven inflation is clarified.  We find
  that these operators do not contribute to the production of
  \emph{Warm Dark Matter} (WDM) and to baryogenesis.  We also
  demonstrate that the sterile neutrino with mass exceeding
  $\unit[100]{keV}$ (a Cold Dark Matter (CDM) candidate) can be
  created during the reheating stage of the Universe in necessary
  amounts.  We argue that the mass of DM sterile neutrino should not
  exceed few MeV in order not to overclose the Universe.}


\renewcommand{\baselinestretch}{1.02}
\begin{document}
\renewcommand{\baselinestretch}{1.1}

\section{Introduction}

The statement that the Universe was dense and hot in the past is an
established experimental fact.  It follows from existence of the
Cosmic Microwave Background radiation (CMB), which has a perfect
Planck spectrum, and from accordance of predictions of the Big Bang
Nucleosynthesis (BBN) with observations.  The latter tells that the
Universe had the temperature of at least few MeV\@.  Whether the
Universe was even hotter is an open question, which hardly can be
answered by experimental means and thus is biased by theoretical
prejudice.

According to the current theory of inflation (for a recent review see
\cite{Linde:2007fr}) the early evolution of the Universe can be
roughly divided into three parts.  During the first, inflationary
stage, the Universe expands exponentially and becomes nearly flat. At
this stage relic gravity waves and matter perturbations, 
leading to structure formation, are generated.  
During the second, reheating stage, the energy stored in
the inflaton field is transferred to the fields of the Standard Model
and other (hypothetical) particles (if they exist).  The third stage
is the radiation dominated Universe in nearly thermal equilibrium for
most of the SM particles.  The starting moment of this stage
$t_r$ corresponds to a maximal temperature of the Universe
$T_\mathrm{max}$, and this is the onset of the standard Hot Big Bang.

The system in thermal equilibrium is completely characterised by
temperature $T$ and chemical potentials $\mu_i$ for exactly conserved
quantum numbers $\mathcal{Q}_i$; the corresponding operators
$\hat{\mathcal{Q}}_i$ obey
$[\hat{\mathcal{Q}}_i,\hat{\mathcal{H}}]=0$, where $\hat{\mathcal{H}}$
is the Hamiltonian of the system.  For the expanding Universe the
precise thermal equilibrium never exists. To describe the state of the
Universe at $T \sim T_\mathrm{max}$ the set of operators
$\hat{\mathcal{Q}}_i$ should be supplemented by approximately
conserved operators $\hat{\mathcal{Q}}_A$, whose rate of change is
much smaller than the rate of the Universe expansion.  Thus, to follow
the Universe evolution at later times, $t > t_r$, one can use the
ordinary kinetic approach based on Boltzmann equations (or equations
for density matrix, if coherent quantum effects are essential) with
initial density matrix
\begin{equation}
  \rho_0 \propto
  \exp\left(
    -\frac{\hat{\mathcal{H}}}{T_\mathrm{max}}
    -\sum_i\frac{\mu_i}{T_\mathrm{max}} \hat{\mathcal{Q}}_i
    -\sum_A\frac{\mu_A}{T_\mathrm{max}} \hat{\mathcal{Q}}_A
  \right)
  \;.
\end{equation}
The magnitude of the maximal temperature $T_\mathrm{max}$ together
with the set of values of the chemical potentials $\mu_i$, $\mu_A$ can
be called \emph{the initial conditions for the Hot Big Bang.}  If they
are known, the further evolution can be completely specified by the
standard methods of kinetic theory.

Clearly, to find the initial conditions for the Big Bang one has to
know what are the relevant particle degrees of freedom at
$T<T_\mathrm{max}$ (in particular, if any new particles beyond those
already present in the SM exist), or, in other words, what is the
Hamiltonian $\hat{\mathcal{H}}$.  The knowledge of the Hamiltonian
would allow to determine the set of conserved $\hat{\mathcal{Q}}_i$ or
nearly conserved $\hat{\mathcal{Q}}_A$ operators and identify the
relevant chemical potentials.  Now, to determine $T_\mathrm{max}$ and
$\mu_{i,A}$ the interaction of the inflaton with the fields in
$\hat{\mathcal{H}}$ must be known, and the physics of reheating must
be elucidated.

Basically, to find the initial conditions for the Big Bang one should
have at hand the theory which is valid up to 
the scale of inflation.  There are quite a number
of proposals for these types of theories, based on different ideas
about physics beyond the SM\@.  These ideas include low energy
supersymmetry and Grand Unification, small, large or infinite extra
dimensions (see e.g.\ \cite{Raby:2008gh} and \cite{Rubakov:2001kp} for
reviews) and many others.  Clearly, any model of physics beyond
the SM, must be able to explain the observed phenomena that cannot be
addressed by the SM physics. They include neutrino masses and
oscillations, the existence of dark matter in the Universe, baryon
asymmetry, inflation, and accelerated expansion of the Universe at
present.\footnote{Perhaps, the observed accelerated expansion of the
  Universe should not necessarily be included in this list as it may be
  irrelevant for the early stages of the Universe evolution we are
  interested in this work.}

The most economical particle physics model which is capable of solving
in a unified way all these problems of the SM is the $\nu$MSM
(Neutrino Minimal Standard Model) of \cite{Asaka:2005an,Asaka:2005pn}.
This theory is nothing but the SM augmented by three relatively light
(lighter than $Z$ boson) right-handed singlet fermions.  If the
dilaton field is added to the $\nu$MSM, the theory can be made
scale-invariant at the quantum level by a specific renormalization
procedure \cite{Shaposhnikov:2008xi,Shaposhnikov:2008ar}.  The
spontaneous breaking of the scale invariance leads then to generation
of all mass parameters, including the Newton's gravity constant.  
Higgs mass is stable against quantum corrections, cosmological
constant is equal to zero, while dark energy, leading to the late
acceleration of the Universe, appears if general relativity is
replaced by the unimodular gravity \cite{Shaposhnikov:2008xb}.
Different phenomenological and cosmological aspects of this theory,
together with the study of how to search for new particles, can be
found in refs.\
\cite{Asaka:2005an,Asaka:2005pn,Bezrukov:2005mx,Boyarsky:2006jm,%
  Asaka:2006ek,Shaposhnikov:2006nn,Asaka:2006rw,%
  Asaka:2006nq,Bezrukov:2006cy,Gorbunov:2007ak,Shaposhnikov:2007nj,%
  Shaposhnikov:2008pf,Laine:2008pg}.  In \cite{Shaposhnikov:2007nj} it
was argued that this model may be valid all the way up to the Planck
scale (for a similar argument in a related theory, see
ref.~\cite{Meissner:2006zh}).  Many of the parameters of this model
are already fixed or constrained by existing cosmological observations
and particle physics experiments.

In \cite{Bezrukov:2007ep} it was found that the Higgs boson of the SM
can play a role of the inflaton, if its non-minimal coupling to the
gravity Ricci scalar is large enough. 
Exactly the same mechanism works in $\nu$MSM and its 
scale-invariant version with 
dilaton \cite{Shaposhnikov:2008xb}.  The evolution of
the dilaton in the latter model with 
phenomenologically interesting choice of parameters 
happens to be irrelevant for inflation.

Reference \cite{Bezrukov:2007ep} provides a rough upper limit on the
maximal temperature of the Universe.  The aim of this work is to
demonstrate that the problem of the initial conditions for the Big
Bang can be solved unambiguously in the SM and in the $\nu$MSM, and to
find these initial conditions.  To this end we consider in detail how
the energy stored in Higgs-inflaton gets transferred to the SM and
$\nu$MSM degrees of freedom.  This allows to make a refined estimate
of the reheat temperature and to fix the concentrations of the singlet
fermions before the hot stage.  We show that the abundances of new
particles are too small to influence the low temperature baryogenesis
in the $\nu$MSM studied in
\cite{Asaka:2005pn,Shaposhnikov:2006nn,Shaposhnikov:2008pf},\footnote{See
\cite{Akhmedov:1998qx} for a suggestion to use singlet fermion
oscillations for leptogenesis.} and low temperature dark matter
production worked out in
\cite{Asaka:2006rw,Asaka:2006nq,Laine:2008pg}.\footnote{For an
original proposal of sterile neutrino as a dark matter candidate see
\cite{Dodelson:1993je,Shi:1998km,Dolgov:2000ew}, earlier computations
of sterile dark matter abundance can be found in
\cite{Dodelson:1993je,Dolgov:2000ew,Abazajian:2001nj,Abazajian:2005gj}.}

The Lagrangian of the SM or of the $\nu$MSM, which can be considered
as the effective theories, can contain all sorts of higher dimensional
operators, suppressed by the Planck mass.  Therefore, we consider the
influence of these operators on inflation and on production of singlet
fermions of the $\nu$MSM.  We find that these operators are definitely
not essential for baryogenesis and for dark matter production, if mass
of the lightest sterile neutrino is below $\unit[100]{keV}$.  In other
words, the conclusion that the production of \emph{WDM} sterile
neutrinos with mass in the keV region must be due to their mixing with
active neutrinos is a robust consequence of the $\nu$MSM\@. Since the
presence of higher-dimensional operators looks to be a generic
phenomenon, we argue that the DM sterile neutrinos have to be lighter
than few MeV in order not to overclose the Universe.

The paper is organized as follows.  In section \ref{sec:Model} we
review the mechanism of inflation based on the Higgs boson of the
Standard Model, and determine the relevant interactions of the
Higgs-inflaton with the other fields of the SM\@.  In section
\ref{sec:reheating} we analyse different processes reheating the
Universe after inflation and estimate the maximal temperature
$T_\mathrm{max}$.  In section \ref{sec:initial-conditions} we discuss
the approximate conservation laws in the SM and the $\nu$MSM and
define the operators $\hat{\mathcal{Q}}_A$.  Then we estimate the
values of chemical potentials $\mu_A$ generated by renormalizable
interactions existing in the $\nu$MSM\@ and the SM\@.  In section
\ref{sec:ster-neutr-prod} we add to the theory higher dimensional
operators and analyse their influence on Higgs-driven inflation and on
the generated values of the chemical potentials.  In section
\ref{sec:hdim-to-baryons} we study the effects of CP-violation at the
reheating stage.  Section \ref{sec:conclusions} contains conclusions.

\section{Higgs-driven inflation}
\label{sec:Model}

The inflationary model with the Higgs boson as the inflaton
\cite{Bezrukov:2007ep,Bezrukov:2008cq} adds the non-minimal coupling
with gravity to the action of the SM (or $\nu$MSM)
\begin{equation}
  \label{Lmain}
  S_J= S_{\mathrm{SM}} + \int d^4x\sqrt{-g}\,\left(
    - \frac{M^2}{2} R -\xi \Phi^\dagger \Phi R
  \right)
  \;.
\end{equation}
Here $S_{\mathrm{SM}}$ is the SM action, $M$ is some mass parameter,
which is nearly equal to the Planck mass in our case, $R$ is the
scalar curvature, $\Phi$ is the Higgs doublet, and $\xi$ is a constant
fixed by the requirement of correct scale of the CMB fluctuations.
Index ``J'' stands for the ``Jordan frame'' action. Action
\eqref{Lmain} contains all  
possible terms of dimension 4 without higher derivatives.\footnote{One
  could also add other dimension 4 terms like $R^2$,
  $R_{\mu\nu}R^{\mu\nu}$, etc., but they lead to terms with higher
  derivatives in the equations of motion and, therefore, lead to
  additional degrees of freedom, which should be dealt with in some
  special way.  We do not consider such extensions here.}  In this
section we review shortly the inflation analysis of
\cite{Bezrukov:2007ep,Bezrukov:2008cq} and introduce some formulas
important for the study of the reheating period.

The only part of the action relevant for inflation is the scalar
sector. In the unitary gauge with
$\Phi(x)=\frac{1}{\sqrt{2}}{0\choose v+h(x)}$ it has the form
\begin{equation}
  \label{eq:SJ}
    S_{J} = \int d^4x \sqrt{-g}\, \Bigg\{
      - \frac{M^2+\xi h^2}{2}R
      + \frac{\dm h\partial^\mu h}{2}
      - \frac{\lambda}{4}\left(h^2-v^2\right)^2
    \Bigg\}
    \;,
\end{equation}
and Higgs vacuum expectation value is $v=\unit[246]{GeV}$.
Another part we analyse later on, while we always stick to the
unitary gauge for simplicity.

\paragraph{The conformal transformation.}
\label{sec:conf-transf}

The simplest way to work with this action is to get rid of the
non-minimal coupling to gravity by making the conformal transformation
from the Jordan frame to the Einstein frame (see, e.g.\
\cite{Kaiser:1994vs,Tsujikawa2000}):
\begin{equation}
  g_{\mu\nu}\to  \hat{g}_{\mu\nu} = \Omega^2 g_{\mu\nu}
  \;,\quad
  \Omega^2 = \frac{M^2+\xi h^2}{M_P^2}
  \;,
\end{equation}
where $M_P\equiv 1/\sqrt{8\pi G_N}=\unit[2.44\times10^{18}]{GeV}$ is
the reduced Planck mass.  This transformation leads to a non-minimal
kinetic term for the Higgs field.  So, it is also convenient to
replace $h$ with new canonically normalised scalar field
$\chi$ by making use of 
\begin{equation}
  \label{eq:3}
  \frac{d\chi}{dh}=\sqrt{\frac{\Omega^2+6\xi^2h^2/M_P^2}{\Omega^4}}
  \;.
\end{equation}
Finally, the action in the Einstein frame is
\begin{equation}
  S_E =\int d^4x\sqrt{-\hat{g}}\, \left\{
    - \frac{M_P^2}{2}\hat{R}
    + \frac{\dm \chi\partial^\mu \chi}{2}
    - U(\chi)
  \right\}
  \;,
\end{equation}
where $\hat{R}$ is calculated using the metric $\hat{g}_{\mu\nu}$ and
the potential is rescaled with the conformal factor
\begin{equation}
  \label{eq:5}
  U(\chi) =
  \frac{1}{\Omega^4\left[ h \l \chi\r \right]} 
  \frac{\lambda}{4}\left[h^2\l \chi\r -v^2\right]^2
  \;.
\end{equation}
We will a bit ambiguously write potential $U$ and scale factor 
$\Omega$ as functions of either $h$ or $\chi$, which should not lead
to misreadings, as far as $h$ and $\chi$ can be expressed one through
another in a unique way.  Figure~\ref{fig:chih} illustrates the
connection between the Higgs field in the Jordan frame, $h$,  and the
Higgs field in the Einstein frame, $\chi$.  
For $\xi\gg1$, the solution of eq.~\eqref{eq:3} can be
approximated in two major regions,\footnote{Exact analytic solution
  exists, but is not really enlightening.} separated by 
\[
\Xcr\equiv \sqrt{\frac{2}{3}}\frac{M_P}{\xi}\;.
\] 
Namely, 
\begin{equation}
  \label{eq:chi(h)}
  \chi \simeq \left\{
    \begin{array}{l@{\qquad\text{for}\quad}l}
      h
      & h<\Xcr
      \;,\\
      \sqrt{\frac{3}{2}}M_P\log \Omega^2(h)
      & \Xcr < h
      \;.
    \end{array}
  \right.
\end{equation}
Note that for analysis of reheating we will need only the field values
smaller than $M_P/\sqrt{\xi}$, where the logarithm in
(\ref{eq:chi(h)}) can be expanded in the following way
\begin{equation}
  \label{eq:chi(h)-reh}
  \chi \simeq
  \sqrt{\frac{3}{2}}\frac{\xi h^2}{M_P}
  \qquad\text{for}\quad
  \Xcr < h \ll \frac{M_P}{\sqrt{\xi}}
  \;.
\end{equation}
Using relations (\ref{eq:chi(h)}) we can explicitly write the
potential as (here we assume $v\ll M_P/\xi$)
\begin{equation}
  \label{eq:U(chi)}
  U(\chi) \simeq \left\{
    \begin{array}{l@{\;\;\;\text{for}\;}l}
      \frac{\lambda}{4} \chi^4
      & \chi< \Xcr
      \,, \\
      \frac{\lambda M_P^4}{4\xi^2}
      \left(\!
        1-\e^{
          -\frac{2\chi}{\sqrt{6}M_P}
        }\!
      \right)^{\!2}
      & \Xcr < \chi
      \,.
    \end{array}
  \right.
\end{equation}
Again, in the region interesting for reheating, the potential can be
approximated by the quadratic potential
\begin{equation}
  \label{eq:U(chi)-quadratic}
  U(\chi) \simeq
  \frac{\omega^2}{2} \chi^2
  \qquad\text{for}\quad
  \Xcr < \chi \ll \sqrt{\frac{3}{2}}M_P
  \;,
\end{equation}
where the ``inflaton mass'' $\omega$ is
\begin{equation}
  \label{eq:omega}
  \omega\equiv\sqrt{\frac{\lambda}{3}}\frac{M_P}{\xi}
  \;.
\end{equation}
Figure~\ref{fig:Ueff} shows schematically the potential
(\ref{eq:U(chi)}).

\DOUBLEFIGURE{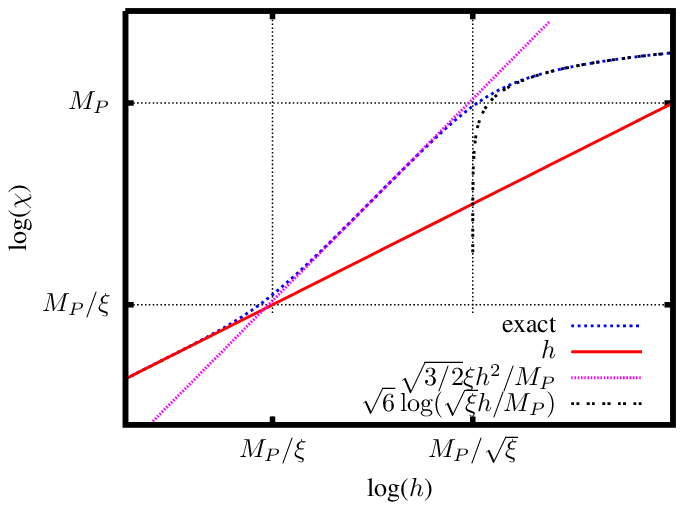}{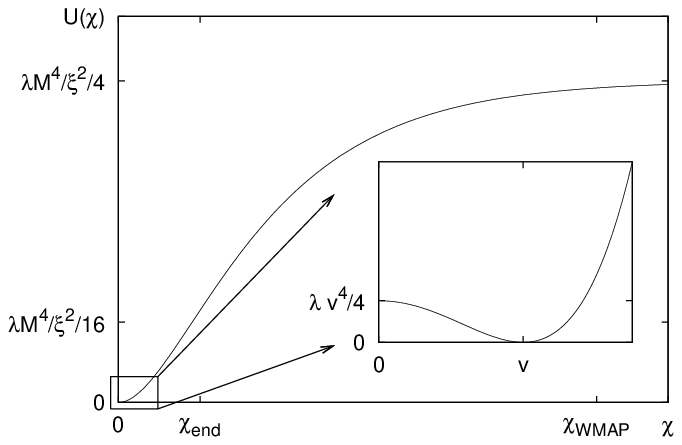}{Dependence of $\chi$ (the Einstein
  frame Higgs field) on $h$ (the Jordan frame Higgs field),
  logarithmic scale.\label{fig:chih}}{Effective potential in the
  Einstein frame.  The insert magnification is not to
  scale.\label{fig:Ueff}}

\paragraph{Inflationary phase.}
\label{sec:inflationary-phase}

The potential \eqref{eq:U(chi)} is exponentially flat for large field
values, and provides the slow roll inflation.  Analysis of the
inflation in the Einstein frame\footnote{The same results can be
  obtained in the Jordan frame
  \cite{Tsujikawa:2004my,Makino1991,Fakir:1992cg}.} can be performed
in the standard way using the slow-roll approximation.  The slow roll
parameters (in notations of \cite{Linde:2007fr}) are easier to express
analytically as functions of the field $h$ using (\ref{eq:3}) and
(\ref{eq:5}), instead of the field $\chi$,
\begin{align}
  \epsilon &
  = \frac{M_P^2}{2}\left(\frac{dU/d\chi}{U}\right)^2
  = \frac{M_P^2}{2}\left(\frac{U'}{U}\frac{1}{\chi'}\right)^2
  \;, \\
  \eta &
  = M_P^2\frac{d^2U/d\chi^2}{U}
  = M_P^2\frac{U'' \chi'-U' \chi''}{U {\chi'}^3}
  \;,
\end{align}
where $'$ denotes derivative with respect to $h$.  Slow roll ends at
$\epsilon\simeq1$, which corresponds to the value $h_\mathrm{end}$.  
The perturbation modes of WMAP \cite{Komatsu:2008hk} 
scale $k/a_0=0.002/\mathrm{Mpc}$ left horizon when the field value
equals $h_\mathrm{WMAP}$. The latter is determined by the number of
inflation e-foldings, 
\begin{equation}
  \label{eq:8}
  N = \int_{h_{\mathrm{end}}}^{h_\mathrm{WMAP}}
  \frac{1}{M_P^2}\frac{U}{U'}\left(\chi'\right)^2dh
  \;.
\end{equation}
To generate proper amplitude of the density perturbations the
potential should satisfy at $h_\mathrm{WMAP}$ the normalization
condition
\begin{equation}
  \label{WMAPnorm}
  U/\epsilon=24\pi^2\Delta_\mathcal{R}^2M_P^4\simeq(0.0276M_P)^4
  \;.
\end{equation}
For
usual quartic potential inflation this condition fixes the coupling
constant $\lambda$, while in our case this allows to find the value
for $\xi$ for \emph{any} given value of $\lambda$.  The inflationary
predictions (see, e.g., \cite{Linde:2007fr}) for the CMB spectrum
parameters are then given by the expressions for spectral index $n_s$
and tensor-to-scalar perturbation ratio $r$,
\begin{equation}
  n_s=1-6\epsilon+2\eta
  \;,\quad
  r=16\epsilon
  \;,
\end{equation}
also calculated at $h_\mathrm{WMAP}$.

In the case of the standard Higgs potential
$V(h)=\frac{\lambda}{4}(h^2-v^2)^2$ we get for the slow roll
parameters \cite{Bezrukov:2008cq} (in the limit
$h^2\gtrsim{}M_P^2/\xi\gg v^2$, $\xi\gg1$, exact expressions can be
found in \cite{Kaiser:1994vs}),
\begin{equation}
  \epsilon \simeq \frac{4 M_P^4 }{3\xi^2h^4}
  \;,\quad
  \eta \simeq
  \frac{4 M_P^4}{3 \xi^2 h^4 }\left(1-\frac{\xi h^2}{M_P^2}\right)
  \;. 
\end{equation}
Inflation ends in at
$h_\mathrm{end}\simeq(4/3)^{1/4}M_P/\sqrt{\xi}\simeq1.07M_P/\sqrt{\xi}$
(and $\chi_\mathrm{end}\simeq 0.94M_P$).  The number of e-foldings is
\eqref{eq:8}
\begin{equation}
  \label{eq:N}
  N =
  \frac{3}{4}\left[
    \frac{h_\mathrm{WMAP}^2-h_{\mathrm{end}}^2}{M_P^2/\xi}
    +\log\frac{1+\xi h_\mathrm{end}^2/M_P^2}
              {1+\xi h_\mathrm{WMAP}^2/M_P^2}
  \right]
  \;,
\end{equation}
leading to
$h_{\mathrm{WMAP}}\simeq9.14M_P/\sqrt{\xi}$.  Thus, the WMAP
normalization (\ref{WMAPnorm}) requires (for $N=59$)
\begin{equation}
  \label{eq:9}
  \xi \simeq 47000\sqrt{\lambda}
  \;,
\end{equation}
where $\lambda$ is the Higgs boson self coupling constant, taken at
\emph{inflationary scale.}  Note, that we retained here the
logarithmic term in (\ref{eq:N}), which was left out in
\cite{Bezrukov:2007ep}.  This, together with WMAP5 value for
normalization (\ref{WMAPnorm}), changed the numerical value in the
relation (\ref{eq:9}).  This does not significantly change the
spectral index and tensor to scalar ratio.

The spectral index is $n_s\simeq1-8(4N+9)/(4N+3)^2$, and the
tensor-to-scalar perturbation ratio is $r\simeq192/(4N+3)^2$.

The number $N$ of e-foldings is fixed from the post-inflation 
history of the Universe described in section \ref{sec:reheating}.  
We show there
that the inflationary stage is followed by the matter dominated epoch,
corresponding to oscillations of Higgs-inflaton with frequency
$\omega$, defined in (\ref{eq:omega}).  The radiation dominated era
starts at effective temperature $T_r$, given by (\ref{eq:56}).  Then,
the number of e-foldings is 
(see \cite{Liddle:1993fq})
\begin{align}
  \notag
  N&=62-\log\frac{k}{a_0H_0}
  -\log\frac{\unit[10^{16}]{GeV}}{U^{1/4}(\chi_\mathrm{WMAP})}
  +\log\frac{U^{1/4}(\chi_\mathrm{WMAP})}{U^{1/4}(\chi_\mathrm{end})}
  -\frac{1}{3}\log\frac{U^{1/4}(\chi_\mathrm{end})}
                       {\rho^{1/4}\l T_{max}\r}
  \\
  &\simeq
  60.4 - \log\frac{k}{a_0H_0} - \frac{1}{6}\log\frac{\Xcr}{X_r}
  \;.
\end{align}
{\sloppy 
Here the present Hubble parameter is $H_0=0.7/(\unit[3000]{Mpc})$, 
$U(\chi_\mathrm{WMAP})\simeq\frac{\lambda M_P^4}{4\xi^2}$, $\rho\l
T_{max}\r$ is the energy density at the beginning of the hot stage, 
$X_r$ is in the range (\ref{eq:53}).  Then, we get
\begin{equation}
  \label{predict}
  N\simeq 59
  \;,\quad
  n_s\simeq 0.97
  \;,\quad
  r \simeq 0.0034
  \;.
\end{equation}
The predicted values are well within one-sigma border of allowed
region of parameter space, see figure~\ref{fig:wmap}.
\FIGURE{
  \includegraphics[width=0.505\textwidth]{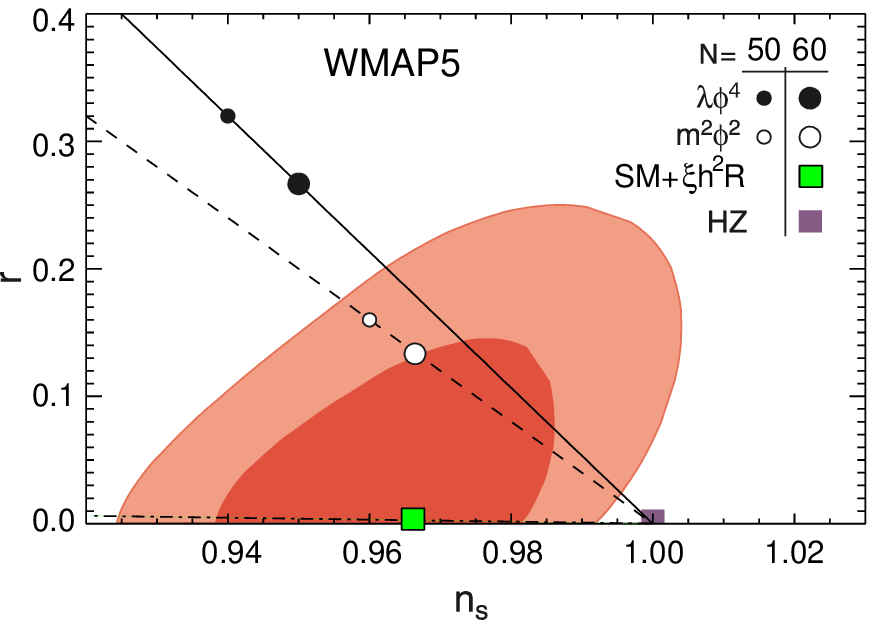}
  \caption{The allowed WMAP+BAO+SN region for inflationary parameters
    ($r$, $n_s$), adopted from \cite{Komatsu:2008hk}.  The green box
    is our predictions supposing 59 e-foldings of inflation.  Black
    and white dots are predictions of usual chaotic inflation with
    $\lambda\phi^4$ and $m^2\phi^2$ potentials, HZ is the
    Harrison-Zeldovich spectrum.}
  \label{fig:wmap}}

\paragraph{Effective couplings in the inflationary domain.}
\label{sec:inflaton-to-SM-couplings}
The inflation and reheating of the Universe occur at energy scales 
much larger, than the electroweak scale. This calls for the study of
radiative corrections to the inflationary potential. A qualitative
discussion of the influence of loop effects on inflation can be found
in \cite{Bezrukov:2007ep}. A number of explicit computations (giving
in some cases conflicting results) has been reported recently 
\cite{Barvinsky:2008ia,Bezrukov:2008ej,DeSimone:2008ei,Bezrukov:2009db,
Barvinsky:2009fy}.  The
conclusion of \cite{Bezrukov:2008ej,DeSimone:2008ei,Bezrukov:2009db,
Barvinsky:2009fy} is that
Higgs-driven inflation is a viable phenomenon in a certain interval of
Higgs masses $m_{\rm min}< m_H < m_{\rm max}$. The values of  $m_{\rm
min}$ and $m_{\rm max}$ found in \cite{Bezrukov:2009db} are:
\begin{align} 
  m_\mathrm{min} & = [126.1 + \frac{m_t - 171.2}{2.1}\times4.1
  -\frac{\alpha_s-0.1176}{0.002}\times 1.5]~\unit{GeV}
  \;,
  \label{mmin}\\
  m_\mathrm{max} & = [193.9
    + \frac{m_t - 171.2}{2.1}\times0.6
    -\frac{\alpha_s-0.1176}{0.002}\times 0.1]~\unit{GeV}
  \;.
  \label{mmax}
\end{align}
with a theoretical uncertainty
$\delta_\mathrm{theor}=\pm\unit[2]{GeV}$.  The similar result for
$m_\mathrm{min}$ was found in an earlier paper \cite{DeSimone:2008ei}
and also in \cite{Barvinsky:2009fy}.

What is essential for the present study of reheating, is the
magnitute of  the coupling constants in the relevant energy domain
$\sim M_P/\xi$. So, an  appropriate renormalization group running of the
coupling constants should be taken into account. Specifically, one
should use the $M_P/\xi$ scale value for the electroweak coupling
constant
\begin{equation}
  \alpha_W^{-1} \simeq 43 \;.
\end{equation}
and the corresponding numbers for the strong and U(1) gauge couplings.

The dependence of the value of the scalar self-coupling  $\lambda$ at
the scale $M_P/\xi$ on the Higgs boson mass  is illustrated in
figure~\ref{fig:lambda} (see ref.~\cite{Bezrukov:2009db} for detailed
description). It can be seen, that out of the window
(\ref{mmin},\ref{mmax}) for Higgs masses  the inflationary scale Higgs
self-interaction starts to behave badly at the energy scale of
inflation. For large Higgs masses it becomes large and thus leads to
strong coupling. For small Higgs masses it gets negative and leads to
instability of the electroweak vacuum. The analysis of the present
paper is not applicable very close to the boundaries of the allowed
region. If the Higgs mass is approaching (\ref{mmin}) or (\ref{mmax})
one has to redo the analysis including higher order radiative
corrections to the Higgs potential, what is beyond the scope of this
paper. 

\FIGURE{\includegraphics{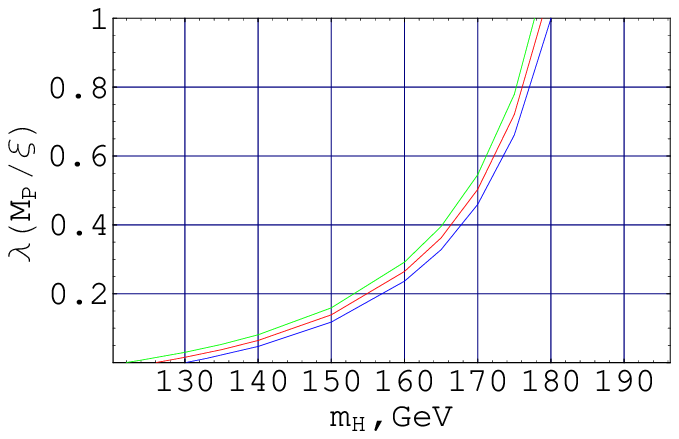}
  \caption{$\lambda$ at the scale $M_P/\xi$ depending on the
  Higgs boson mass $m_H$ for $m_t=\unit[169.1,171.2,173.3]{GeV}$ (from upper
  to lower graph).  Varying e-foldings number and an 
  error in the WMAP normalization measurement one introduces changes
  invisible on the graph.}
  \label{fig:lambda}}

Yet another remark concerns applicability of perturbation theory at
high momenta. As was found in \cite{Sibir-private} and in 
\cite{Burgess:2009ea,Barbon:2009ya},  the perturbation theory, which
is used throughout the calculations of reheating, is inapplicable for
momenta above $M_P/\xi$ in a potential of the form (\ref{eq:5}).
However, at reheating the energy is mostly contained in particles with
momenta $\sim\lambda M_P/\xi$ (see below), which interact weakly.
Thus, the details of the formulation of the theory for high momenta
are irrelevant for present calculations.

\section{Reheating in Higgs-driven inflation}
\label{sec:reheating}

\subsection{Qualitative picture}
\label{sec:overview}
The scalar potential for the Higgs field\footnote{We will use the
names ``Higgs boson'' and ``inflaton'' interchangeably, depending on
the context.}  in the Einstein frame exhibits three qualitatively
different behaviours, leading to three stages of the Universe
expansion.  The first, inflationary stage, corresponding to the flat
potential at $\chi > M_P$, has been already discussed in section
\ref{sec:Model}.  The second specific region of the scalar field
values is
\begin{equation}
  \label{eq:Xcr}
  M_P > \chi > \Xcr = \sqrt{\frac{2}{3}}\frac{M_P}{\xi}
  \;,
\end{equation}
where the scalar potential is essentially \emph{quadratic,} see
(\ref{eq:U(chi)-quadratic}).  The slow roll inflation terminates at
$\chi\sim{}M_P$ with the onset of the oscillations of the scalar
field.  Since the effective inflaton mass $\omega$ is non-zero for
these field values, the exponential expansion of the Universe is
changed to the power low, corresponding to matter domination.  The
amplitude of the Higgs field during this stage is decreased due to
expansion of the Universe and due to particle creation.  At last, for
$\chi<\Xcr$ we are certainly in the radiation-dominated epoch: the
potential for the Higgs field (\ref{eq:U(chi)}) does not contain any
essential mass parameters and thus is scale-invariant; the scalar
self-coupling and couplings of the Higgs field to the fields of the SM
are relatively large, and lead to a rapid energy transfer from the
coherent oscillations to relativistic particles.  Assuming the instant
conversion of the energy of coherent oscillations to relativistic
degrees of freedom of the SM, we get a lower bound on reheating
temperature $T_{\mathrm{reh}}\gtrsim \unit[1.5\times 10^{13}]{GeV}$
(see \cite{Bezrukov:2007ep,Bezrukov:2008cq}).

However, as we show in this section, creation of particles happens to
be important even for $\chi > \Xcr$, and the reheat temperature is
higher.  In what follows we will be mostly interested in the very 
moment, when matter dominated expansion is replaced by the radiation
dominated one. That is when energy in coherent oscillations of the
scalar field is equal to energy collected by SM particles.  We will
characterise this moment by an effective temperature $T_r$ (the ``r''
stands for ``radiation dominance''). It is determined by equating of
the would-be thermally equilibrium energy of SM plasma described by
$T_r$ to its actual energy.  The real thermal equilibrium is achieved
at somewhat lower temperature $T_{\mathrm{reh}} < T_r$.

To determine $T_r$, let us neglect first the effects of particle
creation and consider the evolution of the Universe with the Higgs 
field in the interval (\ref{eq:Xcr}).  The Friedman
equation reads:
\begin{equation}
  \label{eq:freqn}
  H^2(t) = \frac{1}{3M_P^2}\left[
    \frac{\omega^2}{2}\chi^2(t) + \frac{1}{2}\dot\chi^2(t)
  \right]
  \;,
\end{equation}
and leads to a matter dominated expansion regime
\begin{align}
  a &\propto t^{2/3}
  \;,\\
  \chi(t) &= X(t)\cos\left[\omega (t-t_o) \right]
  \label{eq:chi(t)}
  \;,\\
  H(t) &= \frac{\sqrt{\lambda}}{3\sqrt{2}\,\xi} X(t) = \frac{2}{3t}
  \;,&
  X(t) = 2\sqrt{2}\frac{\xi}{\sqrt{\lambda}}\frac{1}{t}
  \label{eq:FriedmannLaw}
  \;.
\end{align}
Here $t$ is the physical time, $a$ is the scale factor, $H$ is the
Hubble parameter, $X(t)$ is the amplitude of the background inflaton
field oscillations, $t_o$ here gives the arbitrary phase of the
oscillations, and $\omega$ is defined in (\ref{eq:omega}).  This
solution is approximate.  It is only reliable for $H\ll\omega$, when
the change of the scale factor is small during one oscillation.  The
amplitude reaches the critical value $\Xcr$ at the critical time
\begin{equation}
  \label{eq:tcr}
  t\approx\tcr\equiv\frac{2\xi}{\omega}
  \;.
\end{equation}

To determine the particle production, we will consider the solution
(\ref{eq:FriedmannLaw}) as an external background.
This approximation breaks down when the energy
of created relativistic particles is comparable with the energy of the
scalar field (inflaton zero mode)
\begin{equation}
  \label{eq:rhoinf}
  \rho_\mathrm{inf}=\frac{\omega^2}{2}X^2=\frac{\lambda}{4}\Xcr^2 X^2
  \;.
\end{equation}
This moment will give us the temperature $T_r$ we are
interested in.


To start with, we describe on the qualitative level various 
processes which occur during the reheating stage at $t<\tcr$ and
single out the most important ones.  Further, we analyse these
processes in detail.

The main mechanism draining energy from the inflaton zero mode is
creation of the particles directly from coupling to the
Higgs-inflaton.  In the background approximation the inflaton field
(\ref{eq:chi(t)}) can be considered as an external source of all 
other fields.  This source has the form of the
varying-with-time masses of all the particles (this includes the
propagating modes of the Higgs field itself).  Only particles with
large couplings to the Higgs field can be created effectively by this
mechanism.  These are the gauge bosons and top quark. Their masses 
in the region \eqref{eq:chi(h)-reh} are
\begin{align}
  m_W^2(\chi) &= \frac{g^2}{2\sqrt{6}} \frac{M_P |\chi|}{\xi}
  \;,\label{eq:mw}\\
  m_t(\chi) &= y_t \sqrt{\frac{M_P|\chi(t)|}{\sqrt{6}\xi}} \sign\chi
  \;.\label{eq:mt}
\end{align}
Here $g^2/4\pi=\alpha_W$ is the weak coupling constant, and
$y_t=\sqrt{2}m_t/v$ is top quark Yukawa.  However, exactly due to large
couplings they are heavy and are still non-relativistic.  The
production of such particles does not 
change the equation of state from the non-relativistic matter to
radiation.  That change happens eventually only due to creation of the
relativistic secondary particles (such as light leptons or quarks)
via  decays or scatterings of the heavy particles.
A competing (but slightly slower) process is the direct creation of
relativistic Higgs excitations, which happens because the potential
\eqref{eq:U(chi)} is not exactly quadratic near the
origin.\footnote{The non-linearity of the potential \eqref{eq:U(chi)}
  at large field values $\chi\sim M_P$ 
is relevant for particle creation only during
  a short period at the very early time, that may be neglected.}

As a result, the generic picture of the reheating process is the
following.  As far as the inflaton ``mass'' $\omega$ is smaller than
the gauge boson \eqref{eq:mw} or top quark mass \eqref{eq:mt} for
$\chi\gtrsim\Xcr$, creation of the gauge bosons or top quarks is
possible only at the moments, when the inflaton field crosses zero
(when $\chi(t)\lesssim\Xcr$).  During each zero crossing some gauge
bosons and top quarks are created.  At first, when the concentration
of the created particles is small (occupation numbers $n_\bk\ll1$),
the creation rate is constant (see Appendix
\ref{sec:nonres-production}).  At this stage the created $W$ bosons
are non-relativistic and decay into light SM fermions (which are
relativistic).  The decay rate, however, changes with time with the
decreasing amplitude of the inflaton oscillations.  The decay process
sustains some quasi constant density of the created bosons (Appendix
\ref{app:sm-boson-decays}).  It stops when the decay rate becomes
smaller than the production rate, which happens at the inflaton
oscillation amplitude \eqref{eq:42}.  Up to this moment no significant
energy transfer from the inflaton to radiation happens.  Then the
generation process accelerates, being enhanced by the stochastic
parametric resonance (occupation numbers $n_\bk>1$, and the
concentration of $W$ bosons rises.  The energy transfer into the light
SM fermions proceeds now mainly via $WW\to f\bar{f}$ annihilation,
(see Appendix \ref{app:sm-boson-scattering}).  This process rapidly
transfers all the energy to radiation, resulting in transition from
the matter domination expansion $a\propto t^{2/3}$ to the radiation
domination $a\propto t^{1/2}$ at field amplitude only slightly smaller
than \eqref{eq:42}.  This should be considered as a refined upper
bound\footnote{More accurate analytical estimate of outcome of the
  scattering processes is hardly possible, since particle masses vary
  quite rapidly.} on critical $X$. The lower bound is given by the
slower energy transfer mechanism~--- the generation of the Higgs bosons on
close-to-vicinity nonlinearities of the potential (\ref{eq:U(chi)})
(see paragraph \ref{app:nr-h-production}), which yields transition at
the oscillation amplitude \eqref{eq:Xtransfer-h2}.  We do not analyse
the production of the top quarks in the present work, because their
contribution is smaller than that of $W$ bosons.  This is because the
parametric resonance enhancement is absent for fermions due to Pauli
exclusion principle.

To summarise, the matter-radiation transition happens when the
amplitude of the inflaton oscillations is somewhere in the region
\begin{equation}
  \label{eq:53}
  3.7 \left(\frac{\lambda}{0.25}\right)^{1/2} \Xcr
  < X_r <
  40 \left(\frac{\lambda}{0.25}\right) \Xcr
  \;.
\end{equation}
The temperature $T_r$ is estimated as follows, 
\begin{equation}
  \label{eq:54}
  g_*\frac{\pi^2}{30}T_{r}^4
  \simeq \frac{\omega^2 X_r^2}{2}
  = \frac{\lambda}{4}\Xcr^2 X_r^2
  \;,
\end{equation}
where $g_*\sim 100$ is the effective number of degrees of freedom of
the SM\@.  This gives for \eqref{eq:53} 
\begin{equation}
  \label{eq:55}
  1.4\times 10^{-5}M_P
  < T_{r} <
  4.5\times10^{-5}  \l \frac{\lambda}{0.25}\r^{1/4}M_P
  \;,
\end{equation}
or
\begin{equation}
  \label{eq:56}
  \unit[3.4\times 10^{13}]{GeV}
  < T_{r} <
  \l \frac{\lambda}{0.25}\r^{1/4} \unit[1.1\times10^{14}]{GeV}  
  \;.
\end{equation}
The coupling constant $\lambda$ here is taken at the inflationary
scale.  Its dependence
on the physical Higgs mass is presented in figure~\ref{fig:lambda}.
Note, that at $T=T_r$ the particle distributions are not yet
fully thermal.

\subsection{$W$ boson production}
\label{subsec:W-production}

Let us start with description of boson production by the
external oscillating source (\ref{eq:chi(t)}).  We write the
equation of motion with the mass given by (\ref{eq:mw}).  Actually,
with formula \eqref{eq:mw} we ignore the exact behaviour
of the mass in time intervals when $\chi<\Xcr$: then the potential is
quartic, so the zero mode evolution also deviates from
(\ref{eq:chi(t)}). Hence, the jump in the derivative in
$m_W^2$ is ``smoothed''.
This introduces an additional cutoff in the
spectrum of produced $W$ bosons, which can be safely neglected. To
simplify the computation, we also replace the vector boson by a
scalar particle.  We will take into account three 
polarizations of the vector boson in the final formula only.

Evolution of the mode $\phi_\bk$ with conformal momentum $\bk$ is
governed by the equation
\begin{equation}
  \label{eq:productioneq}
  \ddot{\phi_\bk}+3H\dot\phi_\bk+
  \left(\frac{\bk^2}{a^2}+m_W^2(t)\right)\phi_\bk = 0
  \;.
\end{equation}
This equation can be solved in the adiabatic
approximation except for the moments when $m_W(t)$ is close to zero.
The reason is that $m_W(t)$ is larger than the frequency
of the background field $\omega$ except for small background
$\chi\lesssim\Xcr$.  The solution of equation \eqref{eq:productioneq} 
in the adiabatic
approximation is (see Appendix \ref{app:semiclassic})
\begin{equation}
  \frac{\phi_\bk}{a^{3/2}} =
  \frac{\alpha_\bk^j}{\sqrt{2k_0}}\e^{
    -i\int_0^tk_0 dt } +
  \frac{\beta_\bk^j}{\sqrt{2k_0}}\e^{
    +i\int_0^tk_0 dt}
  \;,
\end{equation}
where parameters $\alpha_\bk^j$, $\beta_\bk^j$ remain constant between
the moments $t_j$, corresponding to zero background field
$\chi(t_j)=0$.

In the vicinity of the moments $\chi(t_j)=0$ the mass $m_W(t)$ becomes
small compared to the background (source) frequency $\omega$ and
particle creation can take place.  Exact solution in this region (see
Appendix \ref{app:semiclassic}) allows for matching 
$\alpha^{j+1}_\bk,\beta^{j+1}_\bk$ to $\alpha^j_\bk,\beta^j_\bk$.
For the change in the occupation numbers
$n_\bk^j\equiv|\beta_\bk^j|^2$ we have
\begin{equation}
  \label{eq:19}
  \delta n^j_\bk\equiv n_\bk^{j+1}-n_\bk^{j}
  = 
  \frac{|\Rk|^2}{1-|\Rk|^2}
  + \frac{2|\Rk|^2}{1-|\Rk|^2}n_\bk^j
  + \frac{2|\Rk|}{1-|\Rk|^2}\cos\theta_\mathrm{tot}^j
    \sqrt{n_\bk^j(n_\bk^j+1)}
  \;,
\end{equation}
where $\Rk$ is a decreasing with $\bk$ function defined in
(\ref{eq:18'}) (see also figure~\ref{fig:RkDk}), and
$\theta^j_\mathrm{tot}$ is the $\bk$-dependent phase
(\ref{eq:thetatot}).

If the occupation numbers $n_\bk^j$ on the right hand side of
eq.~\eqref{eq:19} are small, $n_\bk^j\ll1$, it describes a simple
particle creation with constant rate.  This is certainly the case if
the particles, generated at each zero crossing, decay before the next
zero crossing, or scatter and change momentum $\bk$ to some value
where the generation coefficients in \eqref{eq:19} are small.  If not,
eq.~\eqref{eq:19} describes a resonance like production of bosons.
There exist several regimes, depending on the model.  The usual
(``narrow'') parametric resonance emerges when $n_j\gg1$ and the phase
$\theta^j_{\mathrm{tot}}/\pi$ changes by an integer number between
$t_j$ and $t_{j+1}$.  In this case one has a pronounced exponential
behaviour for $n_\bk$, located in narrow regions of momentum, which
are defined by 
$\Delta\theta_\mathrm{tot}^j\equiv
\theta_\mathrm{tot}^{j+1}-\theta_\mathrm{tot}^{j}  \simeq n\pi$.  
Another resonance situation
is realized when
\begin{equation}\label{eq:stochresdef}
  \Delta\theta_\mathrm{tot}^j \gg \pi
  \;.
\end{equation}
This regime is called ``stochastic resonance'' \cite{Kofman:1997yn}.
In this case the jumps at moments $t_j$ may be in either direction,
but, on average, they also lead to exponential growth, though a slower
one.  The regime (\ref{eq:stochresdef}) is realised for $X\gtrsim
(\lambda/g^2)\Xcr$ (see (\ref{eq:thetatot}), (\ref{eq:theta})),
which is always true in our case.  In this regime the particle
generation proceeds for any momenta with not too small $|R_\bk|$
(otherwise the last term easily spoils growth of the occupation
number), which implies low momenta.

So, while the number of created particles is small, $n_\bk\ll1$, the
creation of the $W^+$ bosons from (\ref{eq:19}) proceeds in the 
linear regime and can be
approximated as (see Appendix \ref{sec:nonres-production})
\begin{equation}
  \label{eq:linproduction}
  \frac{d(a^3n_{W^+})}{dt}
   \approx 3\cdot a^3 A\frac{\alpha_W}{2\pi^2}\omega^2\Xcr^2 X
  \;,
\end{equation}
with numeric coefficient $A\simeq0.0455$.  The created particles are
essentially non-relativistic.  For concentrations of other gauge
bosons we have the obvious relations $n_{W^+}=n_{W^-}$,
$n_Z=n_{W^+}/\cos^2\theta_W$, where $\theta_W$ is the weak mixing angle.

When the occupation number becomes larger, the production is enhanced
due to the Bose statistics, and is becoming 
approximately exponential (Appendix \ref{app:stochres})
\begin{equation}
  \label{eq:stochres}
  \frac{d(a^3n_W)}{dt}
  \sim a^3 2\omega B n_W
  \;,
\end{equation}
where the numerical coefficient $B\simeq0.045$.  These particles are
also created with nonrelativistic momenta.  Note again, that here the
backreaction of the particles on the condensate is neglected.

\subsection{Transfer into relativistic particles}

Now let us analyse how decay and scattering of the $W$ bosons
influence their generation, described in the section
\ref{subsec:W-production}. 

At small concentrations of the bosons the main process is their decay.
The (average) decay width of the SM gauge boson is
\begin{equation}
  \label{eq:26}
  \Gamma \approx 0.8 \alpha_W \vev{m_W} 
  \;,
\end{equation}
The mass here is
approximated as averaged over inflaton background field oscillations.
Comparing (\ref{eq:26}) with the production rate
(\ref{eq:stochres}), we find that 
for 
\begin{equation}
  \label{eq:42}
  X>\frac{2}{0.64\,\pi}\frac{B^2\lambda}{\alpha_W^3}\Xcr 
  \approx 40 \l \frac{\lambda}{0.25}\r \Xcr
  \;.
\end{equation}
the decay is a more rapid process and prevents
the exponential regime to start.

The energy transfer rate is then balanced by the linear production
(\ref{eq:linproduction}) and the decay rate (\ref{eq:26}).  In
Appendix \ref{app:sm-boson-decays} we show that the energy transfer to
the relativistic modes is negligible for this period.

When the decay process becomes inefficient, the concentration grows,
leading to enhancement of the production approaching exponential
behaviour (\ref{eq:stochres}).  At the same time the main process
responsible for the energy transfer 
to light particles becomes the annihilation of the W bosons,
which is proportional to density squared.  The relevant process is the
annihilation into two fermions $WW\to f\bar{f}$ via t-channel fermion
exchange.  The estimate of the cross section for the process is
\begin{equation}
  \label{eq:47}
  \sigma \approx
  \left(\frac{g}{\sqrt{2}}\right)^4\frac{2N_l+2N_q N_c}{8\pi}
  \frac{1}{\vev{m_W^2}}
  \approx 10\pi\frac{\alpha_W^2}{\vev{m_W^2}}
  \;,
\end{equation}
where $N_l=3$ is the number of lepton generations, $N_q=2+1/4$ is the
effective number of quark generations (virtual $t$-quark contribution
is suppressed by $(m_W/m_t)^2\sim1/4$), and $N_c=3$ is the number of
colours.

Then, the equality between generation and annihilation of the W
bosons is reached at
\begin{equation}
  \label{eq:nscatter}
  n_\mathrm{scatter} = \frac{2B\omega}{\sigma}
  \;.
\end{equation}
The energy drain into the relativistic modes is
\begin{equation}
  \label{eq:49}
  \frac{d}{dt}\left( a^4 \rho \right) =
  2a^4\cdot \sqrt{\vev{m_W^2}} \cdot\sigma n_\mathrm{scatter}^2 
  \;.
\end{equation}
The integral is saturated at late times and gives nearly immediate
transfer of all the energy into the relativistic modes after the
regime (\ref{eq:42}) finishes.

One may note, that this approximation may overestimate $W$ boson
production, and it may actually proceed at a smaller rate, because of
several reasons. First, the occupation number $n_\bk$ is not too large
\eqref{eq:50}, and resonance regime is not fully reached. Second, the
exponent in \eqref{eq:stochres} is actually the upper limit.  Careful
analysis may reveal that the process is slower, so the transfer to the
relativistic degrees of freedom happens later (at lower $X$) than
\eqref{eq:42}, the latter should be considered as the upper bound
on $X_r$.  The lower bound is then given by a slower process of
generation of the Higgs field excitations.

\subsection{Higgs production}

Another particles produced during the inflaton oscillations are
inflaton excitations (Higgs particles).  Really, mass of the
excitations $\delta\chi$ in the background $\chi$ is given by
$m^2_\chi=U''(\chi)$, which is, approximately
\begin{equation}
  \label{eq:mchi}
  m^2_\chi(t) = \left\{
    \begin{array}{l@{\qquad\text{for}\quad}l}
      \omega^2 & \chi(t) > \frac{\omega}{\sqrt{3\lambda}} \;,\\
      3\lambda\chi^2(t) & \chi(t) < \frac{\omega}{\sqrt{3\lambda}}
      \;,
    \end{array}
  \right.
\end{equation}
where $\chi(t)$ is given\footnote{Of course, this is a rough
  approximation for $X\gg \Xcr$, because we should in principle solve
  the equation of motion for $\chi$ in the exact potential.  But at
  large $X\gg\Xcr$ the time, system spent in the region $\chi<\Xcr$,
  is small, and the background solution can be simply approximated by
  \eqref{eq:chi(t)}.} by \eqref{eq:chi(t)}.
Thus, we need to solve the same equation
(\ref{eq:productioneq}), but now with the mass (\ref{eq:mchi}).  The
details of the solution are given in Appendix \ref{app:nr-h-production}.
The produced particles are relativistic, with energy
$E\sim\frac{1}{2}\sqrt{3\lambda}X$, and with energy balance equation 
density
\begin{equation}
  \frac{d(a^3\rho)}{dt}\simeq a^3 \frac{\omega^5}{2\pi^3}\;.
\end{equation}
This competes with the inflaton (zero mode) energy density
$\rho_\mathrm{inf}$ at
\begin{equation}
  \label{eq:Xtransfer-h2}
  X = \frac{M_P}{\xi}\l
    \xi \frac{2\sqrt{6}\lambda}{33\pi^3}
  \r^{1/3}
  \approx 3.7\left(\frac{\lambda}{0.25}\right)^{1/2}
  \left(\frac{\xi}{47000\sqrt{\lambda}}\right)^{1/3} \Xcr
  \;.
\end{equation}
This provides the lower bound on the moment of transition to the
radiation dominated epoch $X_r$.

\subsection{Fermion production from Higgs decay}
\label{sec:fermion-production}

For completeness, let us also calculate the number of light fermions
generated at the reheating stage by the inflaton-Higgs field.  The
result here does not apply to heavy fermions (top quark). The latter
abundance is of small interest: before thermalization it is in any way
not larger than that of the gauge bosons, and after thermalization top
quarks are generated rather fast.

We analyse here the production of the light fermions by the Higgs
condensate decay due to the Yukawa interactions.  The latter are of
the form (\ref{eq:mt}) with some small Yukawa $y$ instead of $y_t$.
The exact treatment would require the solution of  Dirac equation with
the time-dependent mass (\ref{eq:mt}).  To make an order-of-magnitude
estimate, we will replace the source $\propto\sqrt{\sin(\omega t)}$ 
by the simpler one, $\propto\sin(\omega t)$. Though the spectra of the
produced fermions are different (the spectrum is monochromatic for
the sinusoidal source), their total numbers are similar. After this
substitution the fermion time-dependent mass term becomes:   
\begin{equation}
  \label{eq:58}
  y \sqrt{\frac{M_P}{\sqrt{6}\xi X}}X\sin(\omega t) \bar\psi\psi
  \;.
\end{equation}
In the lowest order of perturbation theory the rate of fermion
production here is equivalent to the decay rate  of the system of
scalar particles with mass $\omega$, concentration $n=\omega X^2/2$
and effective Yukawa constant $y \sqrt{\frac{M_P}{\sqrt{6}\xi X}}$:
\begin{equation}
  \label{eq:59}
  \frac{d}{dt}(a^3n_\psi) =
  a^3 \frac{\omega X^2}{2\sqrt{6}}\omega\frac{y^2}{8\pi}\left(
    \frac{M_P}{\xi X}\right)
  \;,
\end{equation}
(and the same formula for the antiparticles $n_{\bar\psi}$). 
An elementary computation  leads to the constant physical particle
density during the matter dominated expansion,
\begin{equation}
  \label{eq:60}
  n_\psi = y^2\frac{\omega^2 M_P}{16\pi}\frac{1}{\sqrt{3\lambda}}
  = y^2 \frac{\sqrt{\lambda}\xi}{32\sqrt{2}\pi}\Xcr^3
  \approx 80 \, \l \frac{\lambda}{0.25}\r y^2\Xcr^3
  \;.
\end{equation}
It is convenient to compare $n_\psi$ with the entropy after the end of
the matter domination stage 
\begin{align}
  s_r &= g_*\frac{4\pi^2}{90}T_r^3
  = g_*\frac{4\pi^2}{90} \left(
    \frac{30\lambda}{4\pi^2g_*}\Xcr^2X_r^2
  \right)^{3/4}
  \notag\\
  &\approx \left[ 2.9\, \l \frac{\lambda}{0.25}\r^{3/2}
    \div
    100\, 
    \l \frac{\lambda}{0.25}\r^{9/4}
  \right]\, \Xcr^3
  \;,
  \label{eq:61}
\end{align}
where we adopted \eqref{eq:55} for the range of $T_r$. 
The resulting abundance is in the range
\begin{equation}
  \label{eq:62}
  \Delta_\psi \equiv \frac{n_\psi}{s_r}
  \approx \left[ 0.8\,\l \frac{\lambda}{0.25}\r^{1/2}
    \div 28\,\l \frac{\lambda}{0.25}\r^{5/4}
  \right]\, y^2
  \;.
\end{equation}
These results are used below for an estimate of primordial
abundance of the sterile neutrinos in the $\nu$MSM\@.

\section{Initial conditions for the hot Big Bang}
\label{sec:initial-conditions}

In section \ref{sec:reheating} we found that when the amplitude of the
Higgs-inflaton drops below $3.8 \Xcr$ the matter-dominated expansion
of the Universe is changed to the radiation dominated behaviour, which
can be characterised at this moment by the effective temperature $T_r
\simeq \unit[3\times 10^{13}]{GeV}$. It is this moment which can be
considered as a starting point for the standard hot Big Bang: the
later evolution of the system can be followed with the use of the SM
or the $\nu$MSM Lagrangian and standard finite temperature equilibrium
and non-equilibrium methods. As we discussed in the Introduction, to
specify the system completely, one has also to determine the values of
the chemical potentials corresponding to either exactly or
approximately conserved quantum numbers. In this section we identify
the most important operators and fix the chemical potentials for them.

Let us start with the SM\@. It has got three anomaly-free exactly
conserved quantum numbers $\mathcal{Q}_\alpha=L_\alpha -\frac{1}{3}B$
($L_\alpha$ is the lepton number of the generation $\alpha$ and $B$ is
the baryonic number). In addition to them, there are quite a number of
approximately conserved different fermionic numbers, such as baryon
number (broken by electroweak anomaly), asymmetry in the number of
right-handed electrons and light quarks such as $u$ and $d$ (broken by
small Yukawa couplings), etc. In the standard inflationary logic one
concludes that all the quantum numbers~--- eigenvalues of corresponding
operators~--- are exponentially small at the end of inflation (at $h
\sim M_P$) and thus can be put to zero.  What concerns the charges
$\mathcal{Q}_\alpha$, they cannot be created in the process of
reheating the Universe, analysed above, simply because they are
exactly conserved. As for the other charges, such as asymmetry in the
light quark flavours, their generation can only occur due to
CP-violating effects. Therefore it is suppressed by the Jarlskog
determinant \cite{Jarlskog:1985ht}, since the only source of
CP-violation in the SM is related of the Kobayashi-Maskawa phase.
Applying the argument of refs.\
\cite{Shaposhnikov:1987tw,Shaposhnikov:1987pf} to this case one
concludes that asymmetries in all CP-odd operators at the beginning of
the Big Bang are at most on the level of $10^{-22}$. To summarize, in
the SM all chemical potentials are negligibly small at the beginning
of the Big Bang. At the same time, the CP-even operators (such as the
abundance of fermions plus antifermions of a given type) equilibrate
with the rate not smaller than $\alpha_W^2 T$, which exceeds the rate
of the Universe expansion right after the beginning of the Big Bang.
So, deviations from thermal equilibrium in the CP-even sector in the
SM can be neglected as well.

Let us turn now to $\nu$MSM\@. The most general renormalizable
Lagrangian containing the SM fields and three right-handed singlet
fermions has the form:
\begin{equation}
  \label{lagr}
  \cL=\cL_{\mathrm{SM}}+
  \bar N_I i \partial_\mu \gamma^\mu N_I
  - y_{\alpha I} \,  \bar L_\alpha N_I \tilde \Phi
  - \frac{M_{IJ}}{2} \; \bar {N_I^c} N_J + \mathrm{h.c.}
  \;,
\end{equation}
where $\cL_{\mathrm{SM}}$ is the Lagrangian of the SM, $y_{\alpha I}$
are new Yukawa couplings, $\Phi$ is the SM Higgs doublet, and
$\tilde\Phi_i = \epsilon_{ij}\Phi_j^*$. In the contrary to the SM,
there are no exactly conserved global quantum numbers in the model.
To analyse the approximately conserved currents, at least the orders
of magnitude of various parameters, entering eq.\ (\ref{lagr}), have
to be fixed. In the $\nu$MSM the Majorana masses of the singlet
fermions are below the electroweak scale and, correspondingly, the new
Yukawa coupling constants are smaller than those in the quark or
charged lepton sector. Their values are constrained by cosmology,
astrophysics and experiment.

The lightest out of the 3 neutral leptons, $N_1$, plays the role of
the dark matter particle.\footnote{We work in the basis in which
  $M_{12}=M_{13}=0$, $M_{23}=M_{32}= M$, $M_1=M_{11} \ll M$,
  $M_{22}\sim M_{33} \ll M$.} Its Yukawa couplings are bounded from
above as
\begin{equation}
  \sum |F_{\alpha 1}|^2 \lsim 10^{-24}
  \label{dm}
\end{equation}
from cosmological considerations \cite{Asaka:2005an,Asaka:2006nq}
related to DM production and from X-ray constraints on the radiative
width of the DM sterile neutrino \cite{Boyarsky:2006jm}. 

The heavier nearly degenerate singlet fermions $N_{2,3}$ (their common
mass is denoted by $M$) fix the pattern of neutrino masses and mixings
and produce baryon asymmetry of the Universe. Their Yukawa couplings
satisfy: \be F_2^2 =\frac{\kappa M m_\mathrm{atm}}{\epsilon v^2} \;,
\ee where $m_\rmi{atm} \approx \unit[0.05]{eV}$ is the atmospheric
neutrino mass difference, $\kappa = 1(2)$ for normal (inverted)
neutrino mass hierarchy, $F_i^2 \equiv [y^\dagger y]_{ii}$, and
$\epsilon = \frac{F_3}{F_2}< 1$. If $\epsilon \sim 1$, i.e.\ for the
case when the couplings of singlet fermions to active leptons are
similar, $F_2^2$ is at most $\sim 2\times 10^{-13}$, corresponding to
$M \sim M_W$. For the smallest possible value of parameter $\epsilon
\simeq 7\times 10^{-5}$ (a lower limit is coming from the requirement
of successful baryogenesis, see figure~10 of \cite{Shaposhnikov:2008pf})
one gets an absolute upper bound on $F_2^2$, \be F_2^2 \lsim 3\times
10^{-11}~,
\label{bau}
\ee
roughly coinciding with the electron Yukawa coupling.

In the limit $F_i \to 0$ the sterile fermions completely decouple from
the fields of the SM, and the $\nu$MSM contains an infinite number of
exactly conserved operators, corresponding to a number of singlet
fermions with any given momentum. Right after inflation these numbers
are exponentially small and can be put to zero. Then, these singlet
fermions are created as described in section 
\ref{sec:fermion-production}.\footnote{Yet another mechanism for
  production of singlet fermions is the decays of $Z$ and $W$ bosons to
  sterile neutrino and left-handed lepton. Since the rate of this
  reaction is suppressed not only by the square of the same Yukawa
  coupling  but also by an extra gauge constant, we expect it to be
  subdominant.}  With the use of (\ref{dm},\ref{bau},\ref{eq:62}) we get
that the abundance of DM sterile neutrinos $N_1$ produced at the
reheating stage is at most $\Delta_1 =n_{N_1}/s\simeq 7\times
10^{-23}$, and the abundance of $N_{2,3}$ is at most $2\times
10^{-9}$. These numbers are too small to play any role in the
subsequent evolution of the Universe. These were the constraints on
CP-even operators, the CP-asymmetries in left-right helicities are 
suppressed much stronger as the CP violating amplitudes must contain
at least two extra powers of Yukawas. 

To summarize, the initial condition for the Big Bang in the $\nu$MSM can
be described by the density matrix
\begin{equation}
  \hat \rho (0) = \hat \rho_\rmi{SM}\otimes |0\rangle\langle 0|
  \;,
  \label{in}
\end{equation}
where $ \hat \rho_\rmi{SM} = Z^{-1}_\rmi{SM} \exp(-\beta \hat{
\mathcal{H}}_\rmi{SM})$, $ \beta \equiv 1/T $, is the equilibrium SM
density matrix at a temperature $T$ with zero chemical potentials, and
$|0\rangle$ is the vacuum state for sterile neutrinos.  The physical
meaning of eq.~\eqref{in} is clear~--- it describes a system with no
sterile neutrinos, while all the SM particles are in thermal
equilibrium. It is this expression which was used for computation of
DM abundance and for computation of baryon asymmetry in the $\nu$MSM
in refs.~\cite{Asaka:2006rw, Asaka:2006nq, Laine:2008pg, Asaka:2005pn,
Shaposhnikov:2008pf}.

\section{Higher dimensional operators}
\label{sec:ster-neutr-prod}

In the first part of the paper we analysed the inflation and reheating
in the model with action \eqref{Lmain}.  However, one may expect that
there are  corrections to this action, suppressed by some large energy
scale (the Planck mass).  We consider the following higher
dimensional operators as an addition to the SM or $\nu$MSM Lagrangian:
\begin{equation}
  \label{eq:10}
  \delta\cL_\mathrm{NR}
 = \frac{\beta}{M_P}\Phi^\dagger \Phi\bar{N}^cN
    -\frac{a_6}{M_P^2}(\Phi^\dagger \Phi)^3
  +\frac{f_{ab}}{M_P}\bar{L_a^c}\Phi \Phi^\dagger L_b
                   +\dots + \mathrm{h.c.}
  \;.
\end{equation}
The natural value of all dimensionless coupling constants is about
one. 

The following questions arise:
\begin{enumerate}
\item Do these operators spoil the picture of inflation discussed
above?
\item Does the reheating change?
\item Can the singlet fermions be created due to these operators in
substantial amounts?
\end{enumerate}

In this section we analyse these issues. Along the lines of
consideration in section \ref{sec:Model}, 
we rewrite first these operators in the  Einstein frame. 
For Higgs part $\delta\cL_{NR}$ this yields a modification of the
potential in (\ref{eq:SJ}), (\ref{eq:5}).  The transformation rule for
the Higgs-fermion interaction is readily obtained, if we also make the
conformal transformation of all the fermionic fields $\psi$
\begin{equation}
  \psi\to\hat{\psi} = \Omega^{-3/2} \psi
  \;.
\end{equation}
The kinetic part for the fermions is conformally invariant, while 
the Yukawa part of the action
\begin{equation}
  \label{eq:1}
    \displaystyle
    S_{J,\mathrm{Yukawa}} =\int d^4x \sqrt{-g} \;
     Y(h)\bar\psi\psi
    \;,
\end{equation}
changes. Here $Y(h)$ describes the generalised Yukawa interaction
providing the fermion $\psi$ with mass. It is not important for the
present discussion whether this is the Majorana or the Dirac
fermion.  The corresponding Einstein frame term is
\begin{equation}
  \label{eq:SE-Yukawa}
    S_{E,\mathrm{Yukawa}} =\int d^4x\sqrt{-\hat{g}} \;
     \frac{Y\left[h\l\chi\r\right]}{\Omega(\chi)} \bar{\hat\psi}\hat\psi
    \;.
\end{equation}
Specifically, for the Dirac mass this yields the Einstein-frame terms
like
\begin{equation}
  S_{E,\mathrm{Dirac}} = \int d^4\sqrt{-\hat{g}}\;
  \frac{m(v)}{v}\frac{h(\chi)}{\Omega(\chi)}\bar{\hat\psi}\hat\psi
  \;,
\end{equation}
which we already used in deriving (\ref{eq:mw}), (\ref{eq:mt}).  Note,
this mass rescaling is similar to that of massive gauge bosons.

For the Majorana higher dimensional term in (\ref{eq:10}) we get
\begin{equation}
  \label{eq:Majormass}
  \delta \cL_{E,\mathrm{NR,Majorana}} =
  \frac{\beta}{2M_P}\frac{h(\chi)^2}{\Omega(\chi)}
  \hat{\bar N}^c\hat N
  \;.
\end{equation}
Note, that at the reheating stage, ($M_P/\xi\ll\chi\ll
M_P/\sqrt{\xi}$), we have $\Omega\sim1$, so the only change is the
field substitution in accordance with (\ref{eq:chi(h)-reh}).

\subsection{Contributions of the higher dimensional operators to
  inflation}
\label{sec:contr-infl-prop}

Clearly, the main effect of the higher dimensional operators is
expected when the Jordan field $h$ is large, of order $M_P$.  Indeed,
at this scale the higher order Higgs field operators may spoil the
flatness of the Einstein frame potential (\ref{eq:U(chi)}), and the
fermion Majorana mass terms can also give sizeable radiative
corrections to the potential. This can change the inflationary
properties of the model. In particular, inflation may turn to be
impossible if the sign of the slope of the potential is changed in the
inflationary region. Or, the predictions of the CMB spectral index and
tensor-to-scalar ratio may leave the experimentally admitted
region. However, the constraints on higher dimensional operators,
imposed by the requirements that the inflation is not spoiled, turn
out to be rather weak. The reason is that the inflationary potential
is only essential at sufficiently small values of the Higgs field,
$h\lesssim h_{WMAP}\sim10M_P/\sqrt{\xi}$. These values are well below
the Planck mass, so that non-ronormalizable contributions are well
suppressed. We analyse below in some detail the contribution of
operators (\ref{eq:10}) to inflationary potential.

\subsubsection{Higgs operators}
\label{sec:Higgs-operators}

Let us analyse the following higher order terms 
added to the Higgs potential: 
\begin{equation}\label{eq:dVn}
  \delta V = \frac{a_n h^n}{2^{n/2}M_P^{n-4}}
  \;,
\end{equation}
$n=6,8,\dots$. As far as the operators are suppressed by the Plank
mass, their effect is mostly important at high values of $h$.  At this
scale we have\footnote{At the end of inflation the values of
  $d\chi/dh$ and $\Omega$ are different, but it only slightly changes
  the WMAP value for $\xi$, as far as the contribution from
  \eqref{eq:dVn} are more suppressed for lower $h$.}
$d\chi/dh\simeq\sqrt{6}M_P/h$, $\Omega^4\simeq \xi^2h^4/M_P^4$.  The
contributions to the slow roll parameters at $N\gsim 60$ are
\begin{align}
  \delta \epsilon &
  = \frac{4(n-4)^2}{3\lambda^2}\frac{a_n^2}{2^n}
  \left(\frac{h}{M_P}\right)^{2n-8}
  \;,\\
  \delta \eta &
  = \frac{2(n-4)^2}{3\lambda}\frac{a_n}{2^{n/2}}
  \left(\frac{h}{M_P}\right)^{n-4}
  \;.
\end{align}
The main contribution comes from the lowest order power term, $h^6$.
Thus, for the change of the parameters at the normalized-to-WMAP 
value of the field we have
\begin{align}
  \delta\epsilon &\sim
  1.7\times10^{-5} a_6^2
  \;,\\
  \delta\eta &
  \sim 0.005 a_6
  \;,
\end{align}
for $\lambda=0.25$.  This implies the change in the spectral index 
$\delta n_s\simeq0.01a_6-0.0001a_6^2$ and in the tensor to scalar
ratio $\delta r\simeq0.0003a_6^2$.

To keep the spectral index within $1\sigma$
bounds $0.94<n_s<0.98$ (at small $r$), see figure~\ref{fig:wmap}, the
coefficient for the dimension six operator in the Higgs potential
should be $|a_6|\lesssim3$.  Hence, no significant
contributions are expected from the higher order operator with
natural values of the coefficients of order one.

\subsubsection{Yukawa terms}
\label{sec:yukawa-terms}

Let us now analyse the effect of the Yukawa terms for the sterile
neutrinos on the inflation. They come from the fermionic loop
contributions to the effective potential for the Higgs field. 
According to \eqref{eq:SE-Yukawa} and \eqref{eq:Majormass} the mass
term for the right handed neutrinos in the Einstein frame has the form
\[
  \cL_\mathrm{mass} = \left[ 
    \frac{M_I}{2\Omega(h)}+ \frac{\beta h^2}{2M_P\Omega(h)}
  \right] \bar{N^c}N
  \;,
\]
where $M_I$ is the usual Majorana mass term for the sterile neutrinos in
$\nu$MSM\@.  For large $h$ the first term is suppressed, but the
second one (dimension 5 operator) provides the mass, growing with the
field.  The latter could change significantly the effective potential
for the Higgs field.  Indeed, the term $yH\bar{L}N$ induces, in the
Einstein frame, the usual Dirac lepton mass, the growth of which stops
at $h\sim M_P/\sqrt{\xi}$, being suppressed by $\Omega(h)$. This
dimension 5 contribution to the mass yields the following contribution
to the Higgs effective potential in the inflation region
\begin{equation}\label{dUeff}
  \delta U(h) = - \frac{m_N^4(h)}{32\pi^2}\log{\frac{m_N^2(h)}{\mu^2}} 
  \simeq
  -\frac{\beta^4h^8}{32\pi^2\Omega^4M_P^4}\log\frac{h^4}{\Omega^2\mu^2}
  \;.
\end{equation}
For high enough $\beta$ this changes the sign of the derivative of
$U(h)$ at some $h$, which would make the inflation impossible or limit
its duration.\footnote{Of course, one could imagine starting inflation
exactly from the top of the potential, where the derivative of $U(h)$
is zero. However, this corresponds to a  highly tuned situation,
keeping in mind that the change of the derivative is due to interplay
between tree level term  and radiative corrections  to the effective
potential.}  Let us calculate the slope of the potential
\begin{equation}
  \frac{dU/d\chi}{U} = \frac{U'}{U\chi'}
  \simeq
  \frac{4}{h\Omega^2} \left(
    1-\frac{h^6\xi^3}{M_P^6}
      \frac{\beta^4\log(h^4/\Omega^2\mu^2)}{8\pi^2\lambda\xi^2}
  \right) \frac{1}{\chi'}
  \;,
\end{equation}
where $'$ means derivative with respect to $h$ and we neglected the
derivative of the logarithm.  It is required that $U'(h)>0$ for at
least 60 e-foldings of inflation, i.e.\ for all $h$ satisfying
$h\sqrt{\xi}/M_P \lsim 10$. The logarithm here, accounting at least
for the inflationary epoch, $1\lsim h\sqrt{\xi}/M_P\lsim10$, is about
$\log(100)\sim 5$. This implies the
constraint
\[
  \frac{\beta^4\log(h^4/\Omega^2\mu^2)}{8\pi^2\lambda\xi^2} <10^{-6}
  \;,
\]
leading to 
\begin{equation}
  \label{eq:betamax}
  \beta^2 \lsim 47 \l \frac{\lambda}{0.25} \r
  \;.
\end{equation}
For smaller Higgs masses the bound is stronger, but always much larger
than one.  We see, that for rather large value of the dimensionless
constant in front of dimension-5 mass operator for the right-handed
neutrinos, the inflation is not spoilt. It is also straightforward to
check by exact calculation of the spectral index that constraints from
the WMAP on $n_s$ lead to essentially the same bound on $\beta$.

\subsection{Sterile neutrino production}

As we have seen in section \ref{sec:fermion-production}, during
preheating the sterile
neutrinos are produced very slowly by
the renormalizable dimension 4 operators.  Let us estimate the
contribution of the dimension-5 operators (\ref{eq:10}) to sterile
neutrino production during and after preheating.  We will separately
analyse the production in the thermal bath \emph{after} reheating
by the annihilation process $hh\to NN$, and \emph{during} 
reheating by the decay of the inflaton-Higgs condensate.

\subsubsection{Thermal production}
\label{sec:thermal-production}

Let us start with the study of neutrino production in the primordial
plasma. In this section we consider the neutrino production
\emph{after} reheating of the Universe, at $T \sim T_r$ (the higher
temperatures are more essential due to the suppression of the relevant
operators by the Planck mass). Here the electroweak symmetry is
restored, and the production of sterile neutrinos goes through
annihilation of the Higgs bosons (4 degrees of freedom, corresponding
to the unbroken phase of the SM) due to coupling (\ref{eq:10}).

The cross section of this process is (neglecting the neutrino mass)
\begin{equation}
  \label{neutrino-production-cross-section}
  \sigma_{hh\to NN }= \frac{\beta^2}{8\pi M^2_P}
  \;. 
\end{equation}
In the absence of other sources of neutrino production, the
interaction \eqref{eq:Majormass} contributes to the r.h.s.\ of the
Boltzmann equation for sterile neutrino density $n_N$
\begin{equation}
  \label{neutrino-production-equation}
\frac{d}{dt}(a^3n_N)=a^34\sigma_{hh\to NN} n_h^2\;.
\end{equation}
Here we took into account the annihilation of all four modes, $n_h$
stands for the density of each scalar degree of freedom.

This equation can be easily integrated accounting for the fact that at
$T<T_r$ the Universe is at the radiation dominated stage:\footnote{We
  suppose that at the temperature $T_r$ the Universe is already in a
  fully thermalised state. Though this is not exactly the case, we
  expect that this assumption can only overestimate the number of
  produced neutrinos, since the non-equilibrium Higgs spectra are more
  enhanced in the infrared region in comparison with the thermal one.}
\begin{gather}
 a\propto \sqrt{t} \;,\qquad
n_h=\frac{\zeta(3)}{\pi^2}T^3
  \;,\\
  H\equiv \frac{\dot a}{a}
=\sqrt{\frac{\pi^2 g_*}{90}}\frac{T^2}{M_P}=\frac{1}{2t}
  \;,\qquad
  s=g_*\frac{4\pi^2}{90}T^3
   \;,
\end{gather}

The solution of equation (\ref{neutrino-production-equation}) gives
the neutrino density-to-entropy ratio,
\[
  \Delta_N \equiv \frac{n_N}{s} =
    \frac{135\zeta^2(3)\sqrt{10}}{4\pi^8g_*^{3/2}} 
    \frac{\beta^2}{ M_P}
    \l T_r-T\r
  \;,
\]
where $g_*=106.75$ is the total number of degrees of freedom of the
SM\@. We also set the initial abundance to zero. We will compute it in
the next chapter, dealing with sterile neutrino production  during 
preheating.

Putting the numbers, we get at low temperatures $T\ll T_r$:
\begin{equation}
  \label{eq:deltaNthermal}
  \Delta_N
  =1.5 \times 10^{-5}\beta^2\frac{T_r}{M_P}
  \;.
\end{equation}

With the use of this relation we can answer the question whether the
primordial thermal production of the lightest practically stable
sterile neutrinos can substantially contribute to the DM abundance.
The neutrino-to-entropy ratio remains intact, so at the current moment
we have
\[
  s_0=\frac{n_{N,0}}{\Delta_N}=\frac{n_{B,0}}{\Delta_B}\;,
\]
where $\Delta_B=0.87\times10^{-10}$ is the baryon-to-entropy ratio and
$n_{B,0}$ is the present baryon number density. Therefore, we can
write for the sterile neutrino abundance $\Omega_N$
\[
\frac{\Omega_N}{\Omega_{DM}} =
\frac{\Omega_B}{\Omega_{DM}}\frac{m_N}{m_p}\frac{\Delta_N}{\Delta_B}\;,
\]
where $m_N$ and $m_p$ are the sterile neutrino and proton masses,
respectively, $\Omega_B=0.046$ and $\Omega_{DM}=0.23$ are  the baryon
and DM abundances  \cite{Komatsu:2008hk}. Hence,
\begin{equation}
  \label{result-thermal}
  \frac{\Omega_N}{\Omega_{DM}}=
  \frac{M_N}{\unit[27]{keV}}\cdot\frac{\beta^2T_r}{M_P}
  \;.
\end{equation}

So, for the maximal allowed $\beta$ (\ref{eq:betamax}) and for the
reheat temperature in the range (\ref{eq:56}), we conclude that the
neutrino mass, required to provide the proper DM abundance, should be
in the range
\begin{equation}
  M_N = \l \frac{0.25}{\lambda}\r\,\left[ 13\, 
    \l\frac{0.25}{\lambda} \r^{1/4}\unit{MeV}\div\unit[42]{MeV}\right]
  \;.
\end{equation}
If  the dimension 5 operator is present with the ``natural''
coefficient $\beta\sim 1$, then the mass of the (long living) sterile
neutrino should not exceed
\begin{equation}
\label{ren-upper-limit}
  M_N<\unit[600]{MeV}\, \l \frac{0.25}{\lambda}\r^{1/4} \;,
\end{equation}
in order not to overproduce the DM (the upper bound from
(\ref{eq:56}) is used for the estimate).

\subsubsection{Production during preheating}
\label{sec:production-during-preheating}

In this section we consider the neutrino production in the early
Universe right after inflation got terminated. The production during
this period happens due to the effective interaction with the Einstein
frame field $\chi$
\begin{equation}
  \label{EffLagrAfterINflation}
  \cL= \l \frac{\beta}{\sqrt{6}\xi}|\chi|\bar N^c N  + \mathrm{h.c.}\r
  \;.
\end{equation} 
We will see that this mechanism produces more sterile neutrinos, than
the thermal production discussed above.

To find particle production due to the time-dependent fermion mass one
has to study the Dirac equation following from
(\ref{EffLagrAfterINflation}). However, to simplify the discussion, we
proceed as in section \ref{sec:fermion-production} and replace
$|\chi|$ by $\chi$. Then the rate of fermion production coincides, to 
the lowest order in Yukawa coupling, with the decay rate of a
collection of scalar particles with certain mass and number density.
An analysis performed in Appendix \ref{app:|h|source} shows that the
number density of produced fermions is not affected by this
replacement, though the spectrum changes.

We can write the Boltzmann equation as
\[
  \frac{d}{dt}(a^3n_N)=a^3\Gamma_{\chi\to NN}n_\chi
  \;,
\]
where we replaced the oscillating source by an ensemble of free scalar
particles with the number density $ n_\chi=\frac{\omega}{2}X^2$ and
particle decay width $\Gamma_{\chi\to
  NN}=\frac{\beta^2}{6\xi^2}\frac{\omega}{8\pi}$.

Then with background  (\ref{eq:chi(t)}), (\ref{eq:FriedmannLaw})
one gets (the early-time contribution is negligible): 
\[
n_N\l X\r = \frac{\beta^2\sqrt{\lambda}}{48\sqrt{2}\pi\xi}\Xcr^2 X
\;.
\]
Dividing this by the entropy (\ref{eq:61}) we get
\begin{align}
  \notag
  \Delta_N &= 
  \frac{\beta^2}{32\pi\sqrt{\pi}} \l
     \frac{30}{\lambda g_*}
  \r^{1/4} 
  \frac{1}{\xi}
  \sqrt{\frac{\Xcr}{ X}}
  =\frac{\beta^2\sqrt{10}}{32\pi^2\sqrt{g_*}\xi^2}\frac{M_P}{T_r}\\
  &= 1.8\times10^{-12}\left(\frac{0.25}{\lambda}\right)
     \frac{\beta^2 M_P}{T_r}
  \;.
\end{align}
This is larger than contribution from thermal generation
(\ref{eq:deltaNthermal}), for the reheating temperature in the range
(\ref{eq:56}).

Proceeding analogously to the previous subsection we get 
\begin{equation}
  \label{result-non-thermal}
  \frac{\Omega_N}{\Omega_{DM}} =
  \beta^2 \frac{M_N}{\unit[2.2\times10^8]{keV}} \l
  \frac{0.25}{\lambda}\r
  \frac{M_P}{T_r}
  \;.
\end{equation}
So, for the maximal allowed $\beta$ given by (\ref{eq:betamax}) and
for the reheat temperature in the range (\ref{eq:56}), we conclude
that the neutrino mass, required to provide proper DM abundance,
should be in the range 
\begin{equation}
  M_N = \unit[65]{keV}\div210\,\l\frac{\lambda}{0.25}\r^{1/4}\unit{keV}
  \;.
\end{equation}
If  $\beta\sim1$, that is what is naturally expected, we have
\[
  M_N<3\,\left(\frac{0.25}{\lambda}\right)~{\rm MeV}
  \;.
\]

Let us summarize the results obtained above. If a more complete, than
the $\nu$MSM, theory leads to higher-order non-renormalizable
operators characterised by a ``natural'' constant $\beta \sim 1$, then
the mass of the DM sterile neutrino must not exceed few
MeV\@.\footnote{For smaller $\beta$ this limit scales as $M_N \propto
  1/\beta^2$.}  Otherwise, it will be produced in amounts enough to
overclose the Universe. Sterile neutrinos, produced at reheating, can
only play the role of CDM, since their mass must exceed
$\unit[65]{keV}$. This requirement comes from the inflationary upper
limit on $\beta$ (\ref{eq:betamax}). Finally, if the sterile neutrino
has a mass in ${\cal O}(10)$~keV region and thus plays a role of WDM
candidate, the thermal primordial production, discussed in this
section, plays no role.

The higher dimensional operators, of course, produce also heavier
singlet fermions of the $\nu$MSM\@. In section
\ref{sec:hdim-to-baryons} 
we analyse whether this has any influence on the 
low-temperature baryogenesis due
to singlet fermion oscillations.

\section{Higher dimensional operators and baryon asymmetry}
\label{sec:hdim-to-baryons}

It is shown in section \ref{sec:ster-neutr-prod} that the abundance of
DM sterile neutrino, created at reheating due to higher dimensional
interactions, cannot exceed
\begin{equation}
  \Delta_N \sim
  10^{-5}
  \;.
  \label{abun}
\end{equation}
Hence, the ``primordial'' (related to inflation) creation of DM is not
effective for light sterile neutrinos, and may play a role only if
$M_N \geq \unit[65]{keV}$. Interestingly, this number is only
somewhat larger than an upper limit on the mass of DM sterile neutrino
produced resonantly due to lepton asymmetry generated in the $\nu$MSM
\cite{Shaposhnikov:2008pf,Laine:2008pg}, $M_1 \lsim \unit[50]{keV}$.
In other words, the initial condition (\ref{in}) is certainly valid
for sufficiently light singlet fermion (mass below $65$ keV), which
could play a role of WDM\@.

Other singlet fermions can be produced due to the same type of higher
dimensional interactions (note that for the abundance computation the
magnitude of the Majorana mass plays no role), and their abundance is
bounded from above by (\ref{abun}). This number is much smaller than
one, meaning that the heavier singlet fermions are practically absent
at the beginning of the hot Big Bang. Still, in order to proof that
the density matrix (\ref{in}) can be used as an initial condition, one
must show that the CP asymmetries in distribution of singlet fermions
do not exceed the baryon asymmetry of the Universe $\sim 10^{-10}$.

It is not difficult to see that this is indeed the case. To this end
consider the most general form of the leptonic part of the Lagrangian,
taking into account the higher dimensional operators of dimensionality
5 as well: 
\begin{equation}
  \label{CP}
  \cL_{CP} = \frac{\beta_{IJ}}{M_P}\Phi^\dagger \Phi\bar{N}^c_I N_J
    + \frac{f_{\alpha\beta}}{M_P}
      \bar{L_\alpha^c}{\tilde\Phi}{\tilde\Phi}^\dagger L_\beta
    - y_{\alpha I} \,  \bar L_\alpha N_I \tilde \Phi + g_{\alpha\beta}
      \bar L_\alpha E_\beta \Phi +\mathrm{h.c.}
  \;,
\end{equation}
where  $E_\beta$ are the right-handed charged leptons.  To get an
amplitude of CP-violating effects, one may consider the imaginary
parts of re-parametrisation invariant products of Yukawa couplings,
which can be considered as a generalization of Jarlskog invariant for
the Kobayashi-Maskawa quark mixing  to this case (see also
\cite{Shaposhnikov:1987tw,Shaposhnikov:1987pf}). These invariants can
be written as traces in flavour space of the products of $\beta_{IJ}$,
$f_{\alpha\beta}$, $y_{\alpha I}$ and $g_{\alpha\beta}$ (no contraction
between Greek and Latin indexes).

The fermion production due to Higgs oscillations and Higgs scattering
appears first to the second order in these couplings. As we have seen
in section \ref{sec:ster-neutr-prod} the leading effect comes from the
first term in (\ref{CP}). Clearly, there is no CP-violation in this
order. To the fourth order in coupling constants the CP-violating
effects appear through the CP-violating trace $\Tr [y^\dagger \beta y
f]$. Since from the flatness of potential $|f| \lsim \beta \lsim 6$,
and because $y_{\alpha I}$ are strongly bounded from above by
(\ref{dm},\ref{bau}), there is a suppression of the asymmetry at least
by 10 orders of magnitude in comparison with (\ref{abun}). Going to
higher orders makes the situation even worse. To conclude, the initial
conditions for the Big Bang are correctly described by eq.~(\ref{in}),
even if higher dimensional operators are included in the $\nu$MSM, and
thus the baryogenesis is a low-temperature phenomenon, having nothing
to do with inflation or Planck scale physics.

\section{Conclusions}
\label{sec:conclusions}

In this paper we analysed in detail the evolution of the Universe in
the scenario where the Higgs boson of the SM plays a role of the
inflaton. The history of the Universe can be divided into three
stages.  The first one is inflation. Here the non-minimal coupling of
Higgs to gravity makes the effective scalar potential flat, the
Universe expands exponentially and the necessary spectrum of
perturbation is generated. This stage finishes roughly at $h \sim
M_P$. During the second stage the Universe expands as under matter
domination. The Higgs field oscillates in the nearly quadratic
potential for $M_P/\xi < h < M_P$, and the particle production is not
effective. When $h$ reaches the critical value $h \simeq M_P/\xi$ the
energy stored in Higgs zero mode is transferred rapidly in other
degrees of the SM, producing the hot Big Bang with temperature
$T_r\approx\unit[10^{14}]{GeV}$.  After this time the Universe
is dominated by radiation.

We have shown that at the onset of the radiation dominated epoch the
densities of all CP-odd operators in the SM can be put to zero and
demonstrated the for the case of the $\nu$MSM the concentrations of
the singlet fermions are negligible at $T_r$. 

We also considered an extension of the SM and $\nu$MSM adding to them
higher dimensional operators suppressed by the Planck scale. We
analysed the constraints on these operators coming from the condition
to have successful inflation. We demonstrated that the concentrations
of the singlet fermions at $T_r$ can be safely put to zero, provided
the mass of DM sterile neutrino does not exceed $\unit[100]{keV}$.
This means that in this case the production of baryon asymmetry and of
dark matter must occur at small temperatures (about and below the
electroweak scale) by essentially the same mechanism, as was described
in \cite{Shaposhnikov:2008pf,Laine:2008pg}. The properties of singlet
fermions can be almost unambiguously fixed by different cosmological
considerations \cite{Shaposhnikov:2008pf,Laine:2008pg}.

We found that the presence of higher-dimensional operators provides a
new mechanism for primordial production of DM sterile neutrino. This
mechanism is effective in models with sufficiently heavy sterile
neutrinos, $M_N \gsim \unit[100]{keV}$.  No presently available
astrophysical constraints (in particular, those associated with
X-rays) can exclude this possibility, since production occurs even if
DM sterile neutrino Yukawa couplings are identically equal to zero.
However, if the solution of the short scale difficulties of the CDM
scenario \cite{Moore:1999nt,Klypin:1999uc,Sommer-Larsen:1999jx,%
  Bode:2000gq,Goerdt:2006rw,Gilmore:2006iy,Strigari:2008ib} is to be
given by the WDM, this region of the parameter space should be
discarded.\footnote{See, however, ref.~\cite{Gorbunov:2008ka}, where it
  is shown that heavy sterile neutrinos could be WDM for other types
  of production mechanisms.}  At the same time, there are no reasons
to expect that these operators are suppressed by the scale exceeding
the Planck one (i.e.\ it is unlikely that $\beta < 1$).  Therefore, the
models with DM sterile neutrinos heavier than $\unit[3]{MeV}$ are
generally disfavoured due to problems with dark matter overproduction.
These arguments provide an extra justification of sub MeV mass of the
lightest singlet fermion within the $\nu$MSM.

\section*{Note added}
\label{sec:note}

Some time after our paper was posted at arXiv the article
\cite{GarciaBellido:2008ab} devoted to the same subject appeared. 
Most of the conclusions of \cite{GarciaBellido:2008ab} are similar to
ours. In particular, the authors of \cite{GarciaBellido:2008ab} used a
similar formalism to analyze the transfer of energy from the Higgs
field oscillations to gauge bosons. A detailed analysis of differences
and similarities of these works goes beyond the scope of the present
paper. We would like just to mention that some (not inessential for
physical consequences) differences in numerics are presumably due to
the fact that annihilation of created gauge bosons was not accounted
for in \cite{GarciaBellido:2008ab}, leading to the different rate of
transfer of energy to relativistic particles at later stages of
reheating.

\acknowledgments

The authors thank S. Sibiryakov and I. Tkachev for valuable comments.
The work of F.B. and M.S. was supported in part by the Swiss National
Science Foundation. The work of D.G. was supported in part by the
grants of the President of the Russian Federation NS-1616.2008.2 and
MK-1957.2008.2, by the RFBR grant 08-02-00473-a and by the Russian
Science Support Foundation. D.G. thanks ITTP EPFL for kind
hospitality.

\appendix

\section{$W$ boson production}

\subsection{Semiclassic approach}
\label{app:semiclassic}

The discussion in this section closely follows \cite{Kofman:1997yn}.
At first approximation a creation of $W$ bosons can be regarded as a
creation of particles with mass (\ref{eq:mw}) varying with the
amplitude of the Higgs field (\ref{eq:chi(t)}).  This approximation
breaks when important amount of energy is transferred from the
inflaton zero mode (\ref{eq:rhoinf}).  If at this moment the energy is
in the relativistic modes this corresponds to the moment $T_r$ of
transition to the radiation dominated expansion.

To solve the equation (\ref{eq:productioneq}) we rescale the variables
by
\begin{equation}
  \varPhi_\bk = a^{3/2}\phi_\bk \;.
\end{equation}
This leads to the equation for an oscillator with varying frequency
\begin{gather}
  \label{eq:11}
  \ddot\varPhi_\bk+k_0^2(t)\varPhi_\bk = 0
  \;,\\
  k_0^2(t)=\frac{\bk^2}{a^2}+m_W^2(t)+\Delta
  \;,
\end{gather}
where $\Delta\equiv-\frac{3}{4} \left( \frac{\dot{a}}{a}\right) -
\frac{3}{2} \frac{\ddot{a}}{a}$ is always small and can be neglected. 
The initial conditions, corresponding to vacuum oscillations, are
\begin{equation}
  \label{eq:12}
  \varPhi_\bk(t) = \frac{\e^{-ik_0 t}}{\sqrt{2k_0}}
  \;.
\end{equation}
Equation \eqref{eq:11} can be solved in the adiabatic approximation
when $\dot k_0\ll k_0^2$.  At $\bk=0$ this condition is equivalent
(up to a change of the scale factor $a$, which is negligible in our
case) to $\dot m_W\ll m_W^2$.  This is true for
\begin{equation}
  \label{eq:30}
  |t-t_j| \gg \left(\frac{\sqrt{6}\xi}{2g^2M_PX\omega}\right)^{1/3}
   = \frac{1}{4^{1/3}K}
   \;,
\end{equation}
where $t_j$ are moments when inflaton crosses zero, so that
$m^2(t_j)=0$, and $K$ is the natural scaling parameter defined in
\eqref{eq:16}.  In these regions adiabatic solution is
\begin{equation}
  \label{eq:13}
  \varPhi_\bk = \frac{\alpha_\bk^j}{\sqrt{2k_0}}\e^{
    -i\int_0^tk_0 dt } +
  \frac{\beta_\bk^j}{\sqrt{2k_0}}\e^{
    +i\int_0^tk_0 dt}
  \;,
\end{equation}
where parameters $\alpha_\bk^j$, $\beta_\bk^j$ remain
constant within $t_{j-1}<t<t_j$.
At the moments $t_j$ the coefficients get changed by the 
Bogolubov transformation
\begin{equation}
  \label{eq:14}
  \begin{pmatrix}
    \alpha_\bk^{j+1}\e^{-i\theta_\bk^j} \\
    \beta_\bk^{j+1}\e^{+i\theta_\bk^j}
  \end{pmatrix}
  =
  \begin{pmatrix}
    \nicefrac{1}{D_\bk}     & \nicefrac{R_\bk^*}{D_\bk^*} \\
    \nicefrac{R_\bk}{D_\bk} & \nicefrac{1}{D_\bk^*}
  \end{pmatrix}
  \begin{pmatrix}
    \alpha_\bk^{j}\e^{-i\theta_\bk^j} \\
    \beta_\bk^{j}\e^{+i\theta_\bk^j}
  \end{pmatrix}
  \;,
\end{equation}
where $R_\bk$ and $D_\bk$ are the ``reflection'' and ``transition''
coefficients for each interval $t_{j-1}<t<t_j$ 
(they obey the equality $|R_\kappa|^2+|D_\kappa|^2=1$),
and
\begin{equation}\label{eq:theta}
  \theta_\bk^j\equiv\int_0^{t_j}k_0 dt
  \;.
\end{equation}
To find the coefficients $R_\bk$ and $D_\bk$ for each interval we need
to solve exactly eq.~\eqref{eq:11} in the vicinities of the moments
when $m_W\simeq0$ (and where the adiabatic approximation is
inapplicable) and match this solution with \eqref{eq:13} in the
intermediate region of the field amplitude.  Obviously, exact solution
is impossible, but we can approximate $m_W^2(t)$ near zero as
$\const\cdot |t-t_j|$.  This approximation to the potential is good
enough at
\begin{equation}
  \label{eq:31}
  |t-t_j|\ll\frac{\sqrt{6}}{\omega}
  \;.
\end{equation}
The regions \eqref{eq:30} and \eqref{eq:31}
intersect for
\begin{equation}
  \label{eq:38}
  X>\frac{\lambda}{48\sqrt{6}\pi\alpha_W}\Xcr
  \;,
\end{equation}
which covers all the possibly interesting reheating period.  At the
same time, one can estimate, that the effects of ``smoothing'' of the
$|t-t_j|$ are also insignificant up to approximately $\Xcr$.  Thus, we
can match the solution of the linear equation and adiabatic solution.
To solve the ``linearised'' equation we rescale the variables as
\begin{equation}
  \label{eq:15}
  \kappa = \frac{|\bk|}{Ka}
  \;,\quad
  \tau=K(t-t_j)
\end{equation}
with
\begin{equation}
  \label{eq:16}
  K\equiv \left[\!
    \frac{g^2M_P^2}{6\xi^2} \sqrt{\frac{\lambda}{2}} X(t_j)
  \!\right]^{\!1/3}\!\!
  =
  \left[
    \omega \frac{g^2}{2\sqrt{6}}\frac{M_P X(t_j)}{\xi}
  \!\right]^{\!1/3}
  \!\!\!\!\!\!\!.
\end{equation}
Then, for small $\tau$ eq.~\eqref{eq:11} takes the form
\begin{equation}
  \label{eq:17}
  \frac{d^2 \varPhi_\bk}{d\tau^2}+(\kappa^2+|\tau|)\varPhi_\bk=0
  \;,
\end{equation}
which can be readily solved analytically in terms of the Airy
functions.  Matching the solution with the asymptotic form
\eqref{eq:13} at $t=t_j$ one gets (see Appendix
\ref{app:modulebarrier} for details)
\begin{align}
  \label{eq:18}
  D_\bk &= \e^{2i\left(\frac{2}{3}\kappa^3+\frac{\pi}{4}\right)}
    \frac{i\left[\Ai'(-\kappa^2)\Bi(-\kappa^2)-\Ai(-\kappa^2) 
\Bi'(-\kappa^2)\right]}
    {(\Bi(-\kappa^2)+i\Ai(-\kappa^2))(\Bi'(-\kappa^2)+i\Ai'(-\kappa^2))}
    \;,\\
  \label{eq:18'}
  R_\bk &= \e^{2i\left(\frac{2}{3}\kappa^3+\frac{\pi}{4}\right)}
    \frac{-\Ai'(-\kappa^2)\Ai(-\kappa^2)-\Bi(-\kappa^2)\Bi'(-\kappa^2)}
         { (\Bi(-\kappa^2)+i\Ai(-\kappa^2))
           (\Bi'(-\kappa^2)+i\Ai'(-\kappa^2)) }
  \;.
\end{align}
Thus, we can calculate the occupation number
$n_\bk^j\equiv|\beta_\bk^j|^2$ at the moment $t_j$ 
\begin{gather}
  \label{eq:niterative}
  n_\bk^{j+1} = \frac{|R_\bk|^2}{|D_\bk|^2} +
  \frac{1+|R_\bk|^2}{|D_\bk|^2}n_\bk^j +
  2\sqrt{1+n_\bk^j}\sqrt{n_\bk^j}\frac{|R_\bk|}{|D_\bk|^2}
    \cos(\theta_{\mathrm{tot}}^j)
  \;,\\
  \label{eq:thetatot}
  \theta_{\mathrm{tot}}^j =
  -2\theta_\bk^j-2\l \frac{2}{3}\kappa^3+\frac{\pi}{4}\r
  +\arg\alpha_\bk^j-\arg\beta_\bk^j
  \;.
\end{gather}

\FIGURE{%
  \includegraphics[width=0.5\textwidth]{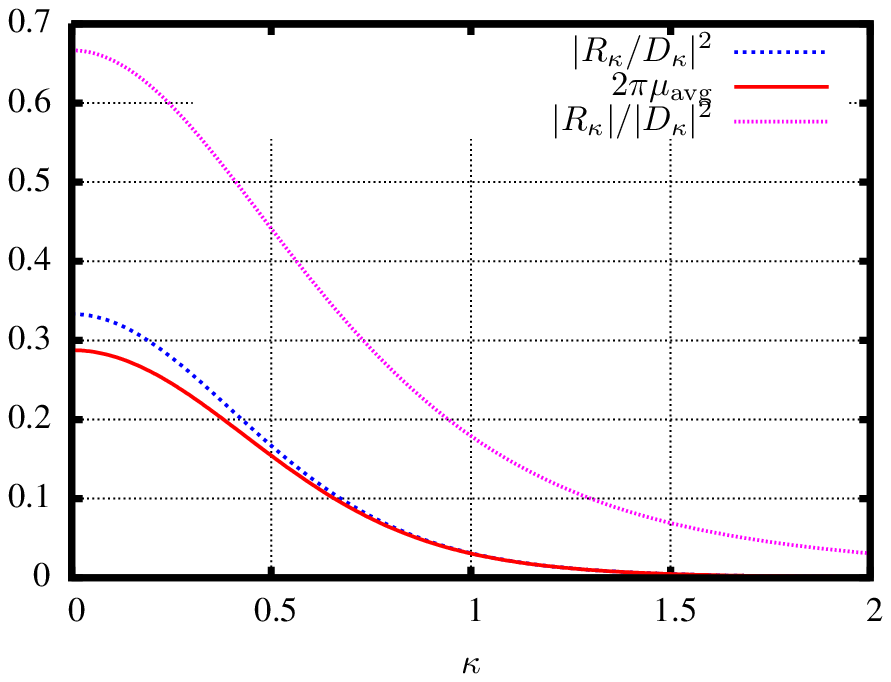}%
  \includegraphics[width=0.5\textwidth]{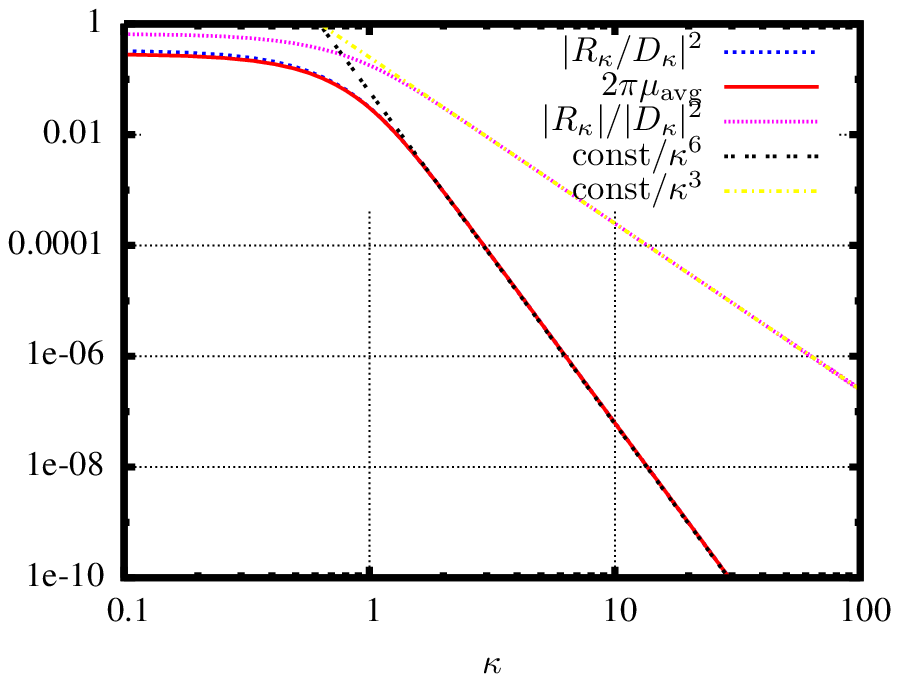}%
  \caption{Particle creation coefficients for (\protect\ref{eq:19}),
    and the effective resonance exponent parameter
    (\protect\ref{eq:41}).  The right plot is in the log-log scale, to
    show $\kappa^{-6}$ and $\kappa^{-3}$ behaviour at large momenta.
    \label{fig:RkDk}}
}

The total particle number density for $t_j<t<t_{j+1}$ is given by
\begin{equation}
  \label{eq:6}
  n(t_j<t<t_{j+1}) = \int\frac{d^3\bk}{(2\pi a)^3} n_\bk^j
  \;.
\end{equation}

\subsection{Non-resonance production}
\label{sec:nonres-production}

In this case, we estimate the production of the particles simply by
the first term in \eqref{eq:niterative} (only one degree of freedom of
$W$ boson is accounted for)
\begin{equation}
  \frac{dn}{dt}=\frac{\omega}{\pi} \int \frac{d^3\bk}{(2\pi)^3}
    \frac{|R_\bk|^2}{|D_\bk|^2} 
   \approx \frac{A}{2\pi^3}\omega K^3
  \;,
\end{equation}
with the numeric coefficient $A=0.0455$.  The spectrum of the created
particles is presented in figure\ \ref{fig:RkDk}.  It has cutoff at
$|\bk|/a\sim K$ and power law tail $|\bk|^{-6}$.  Thus, at $X>\Xcr$
the produced particles are \emph{nonrelativistic,} as far as
$K<\sqrt{\vev{m_W^2}}$.  So, to study the transition to the radiation
domination one should analyse further the $W$ boson decays into
relativistic particles, which is done in the next subsection.

Let us also note, that taking into account three polarizations of the
vector bosons and their different types is made as follows
\begin{equation}
  \label{eq:nWZ}
  n_{W^+}=n_{W^-}=3n
  \;,\quad
  n_Z=\frac{3}{\cos^2\theta_W}n
  \;,
\end{equation}
where $\theta_W$ is the weak mixing angle, and $n_{W^\pm}$, $n_Z$ are
densities for W$^\pm$ and Z-bosons, respectively.

\subsection{Stochastic resonance}
\label{app:stochres}

If the occupation numbers $n_\bk$ exceed one, then the first
term in \eqref{eq:niterative} can be neglected, and the last two
terms, proportional to $n_\bk$, yield the resonance~--- an
exponentially rapid particle creation.  
If $n_\bk\gg1$ and the shift in the phase, 
$\Delta\theta$, is large, then we can neglect the first term in
\eqref{eq:niterative} and write approximately
\begin{equation}
  \label{eq:20}
  n_\bk^{j+1}\simeq n_\bk^j\e^{2\pi\mu_{\bk\,\mathrm{avg}}^j}
  \;.
\end{equation}
The average growth exponent $\mu_{\bk\,\mathrm{avg}}$ is obtained by
averaging of the exponent over the random phase
\begin{equation}
  \label{eq:21}
  \mu_{\bk\,\mathrm{avg}}^j \equiv
  \mu_{\mathrm{avg}}\left(\frac{|\bk|}{K(t_j)a(t_j)}\right) =
  \int_0^{2\pi}
  \frac{d\theta}{2\pi}
  \frac{1}{2\pi}\log\left(
    \frac{1+|R_\bk|^2}{|D_\bk|^2} +
    2\frac{|R_\bk|}{|D_\bk|^2}\cos(\theta)
  \right)
  \;,
\end{equation}
The integral can be found exactly, leading to
\begin{equation}
  \label{eq:41}
  \mu_{\mathrm{avg}}(\kappa) = \frac{-\log(|D_\bk|^2)}{2\pi}
  \;,
\end{equation}
which is presented in figure~\ref{fig:RkDk}.

If we neglect the expansion of the Universe, then the time derivative
of the total particle number can be estimated as
\begin{equation}
  \frac{dn}{dt} = \int\frac{d^3\bk}{(2\pi)^3a^3}
  n_\bk 2\omega \mu_\mathrm{avg}\left(\frac{|\bk|}{Ka}\right)
  \sim 2\omega B n
  \;,
\end{equation}
where the numerical coefficient $B\sim\mu_\mathrm{avg}(0)\simeq0.045$.
The exact expression depends on the exact spectrum of the generated
particles, and can be omitted at our level of precision.  Note, that
the same equation, describing the exponential creation, is true
without any change for $n_{W^\pm}$ and $n_Z$.  The difference appears
in the pre-exponential behaviour only.

Typical momentum of the produced particles is again $K$, so they are
non-relativistic, and analysis of their conversion into light
relativistic particle is needed.  It proceeds via annihilation
(scattering) and is studied in detail in Appendix
\ref{app:sm-boson-scattering}.

\section{$W$ boson decays and scatterings: energy transfer into
  relativistic particles}

\subsection{$W$ boson decays}
\label{app:sm-boson-decays}

Now let us analyse whether some processes, like decay or scattering of
the $W$ bosons, may destroy the resonance picture described in Appendix
\ref{app:semiclassic}.  These processes may
destroy the resonance behaviour by taking the bosons out of the
resonance region in two ways: either by transferring the energy to
other particles, or by changing the boson momenta and taking it out
of the resonance region.

Changing momentum could be expected in $WW\to WW$
scatterings. However, as the typical $W$ boson momentum is smaller
than their mass $K<\vev{m_W}$, one can not achieve in scatterings
momenta larger than $K$, and this process (though rather effective)
can be safely neglected.

So, the two remaining processes are decay and annihilation of the $W$
bosons, which transfer the energy to the relativistic (light)
particles, and, depending on their rate, may also prevent the
development of parametric resonance.

We start with the analysis of the decay process of the gauge bosons
created at the moment $t_j$.  The (average) decay width of the SM
gauge boson is given by (\ref{eq:26}).  We also estimate the $W$ boson
mass as the averaged value over inflaton oscillations,
\begin{equation}
  \label{eq:mWavg}
  \vev{m_W^2}=\frac{g^2}{2\sqrt{6}}\frac{M_P\vev{|\chi|}}{\xi}
  = 2\alpha_W \Xcr X
  \;,
\end{equation}
where $\vev{|\chi|}=\frac{2X(t)}{\pi}$.  If the decay is
faster than the exponent of the stochastic parametric resonance
$2\omega B$ (see \eqref{eq:stochres}), then the parametric resonance
never settles, and creation is dominated by the first term in
\eqref{eq:niterative}, see section \ref{sec:nonres-production}.  The
inequality $\Gamma>2\omega B$ leads to (\ref{eq:42}).  So, for the
period before \eqref{eq:42} the production happens only due to the
first term in \eqref{eq:19}.  Let us check, that the energy in the
decay products of the $W$ bosons remains small for this period.

During the time period \eqref{eq:42}, when the decay is fast, the
creation of the particles is non-resonant \eqref{eq:linproduction}.
We can write the approximate Boltzmann equation for this period 
\begin{equation}
  \label{eq:43}
  \frac{d}{dt}\left( a^3 n \right)
  =
  a^3 \left(
    \frac{A}{2\pi^3}\omega K^3
    -\Gamma n
  \right)
  \;.
\end{equation}
The solution to this equation in the semi-stationary regime,
corresponding to vanishing time derivative in the left-hand side, is
\begin{equation}
  \label{eq:44}
  n_\mathrm{decay} \simeq \frac{A\omega K^3}{2\pi^3\Gamma}
  \;.
\end{equation}
The semi-stationary approximation $|\dot n|,\,3Hn \ll \Gamma n$, is
valid for
$X<\frac{4\alpha_W}{\lambda}(0.8\alpha_W)^2\xi^2\Xcr\approx0.7\times10^5\Xcr$,
that is always after the end of inflation.

We can also check, that the occupation numbers $n_\bk$ are really much
smaller than one and we are in the non-resonant regime
\eqref{eq:linproduction}.  As far as the typical physical momenta of
the $W$ bosons are of the order of $K$ we have
\begin{equation}
  \label{eq:48}
  n_\bk \sim \frac{n_\mathrm{decay}}{K^3}
  =
  \frac{\sqrt{\Xcr}\,A\,\sqrt{\lambda}}{4\,\pi^3\,\sqrt{
        \alpha_W}\,(0.8\alpha_W)\,\sqrt{X}}
  \approx 0.06 \, \l \frac{\lambda}{0.25} \r^{1/2}\,
            \sqrt{\frac{\Xcr}{X}}
  \;,
\end{equation}
which is much smaller than one for all interesting $X$.

The energy during this stage is converted to the SM particles 
produced in $W$ boson decays.  They are light and
relativistic, and typical energy transferred to them in each decay is
of the order $\sqrt{\vev{m_W^2}}$, as far as the $W$ bosons are
non-relativistic.  Thus, we can write the Boltzmann equation in the
expanding Universe for the energy density in the relativistic SM
particles as
\begin{align}
  \label{eq:45}
  \frac{d}{dt}\left( a^4 \rho \right) & \simeq
  a^4 \left(6+\frac{3}{\cos^3\theta_W}\right)
  \sqrt{\vev{m_W^2}}n_\mathrm{decay}\Gamma
  \\\notag
  & \simeq
  a^4 \left(6+\frac{3}{\cos^3\theta_W}\right)
  \sqrt{\vev{m_W^2}} \frac{A\omega K^3}{2\pi^3}
  \;,
\end{align}
where the coefficient $\left(6+\frac{3}{\cos^3\theta_W}\right)$
accounts both for the different number and masses of created W$^\pm$-
and Z-bosons (see eq.~(\ref{eq:nWZ})).  The solution is saturated by
late time and reads
\begin{equation}
  \label{eq:46}
  \rho = \left(6+\frac{3}{\cos^3\theta_W}\right)
  \sqrt{\vev{m_W^2}} \frac{A\omega K^3}{2\pi^3} \frac{6}{13}t
  \;.
\end{equation}
This reaches the inflaton energy density $\rho\sim\rho_\mathrm{inf}$
\eqref{eq:rhoinf} at
\begin{equation}
  \label{eq:decay-equipartition}
  X \simeq \left(6+\frac{3}{\cos^3\theta_W}\right)^{2/3}
  \frac{4\cdot3^{\frac{2}{3}}\,\xi^{\frac{2}{3}}\,\Xcr\,A^{\frac{2
          }{3}}\,\alpha_W}{13^{\frac{2}{3}}\,\pi^{\frac{4
          }{3}}\,\lambda^{\frac{1}{3}}}
  \approx 5.8\Xcr
  \;,
\end{equation}
that is much later than the end of the non-resonant creation period
\eqref{eq:42}.  We conclude, that the energy drain by $W$ boson decays
is irrelevant during the non-resonant inflaton decay.

\subsection{$W$ bosons annihilation}
\label{app:sm-boson-scattering}

Self scattering of the $W$ bosons, like $WW\to WW$ is of little
interest for us, as far as it does not take the bosons out of the
stochastic resonance zone (the bosons are non-relativistic, so after
scattering they retain their small momenta).

It is easy to check, that for the $W$ boson number density
\eqref{eq:44} saturated by the boson decays (discussed in Appendix
\ref{app:sm-boson-decays}),
during the period \eqref{eq:42} of non-resonant inflaton decay the
annihilation to fermions is 
negligible, $\sigma n_\mathrm{decay}^2 < \Gamma
n_\mathrm{decay}$ (relation formally holds for $X>0.1\Xcr$).

The scattering proceeds much more actively at larger particle
densities, so the relevant production mechanism is given by the
stochastic resonance \eqref{eq:stochres}.  Thus, we can 
approximate the effective Boltzmann equation for the $W$ boson
particle number as
\begin{equation}
  \label{eq:beq-scatter}
  \frac{d}{dt}\left( a^3 n_{W} \right)
  =
  a^3 \left(
    2B\omega n_{W}
    -\sigma n_{W}^2
  \right)
  \;,
\end{equation}
where $n_W=n_{W^+}=n_{W^-}$.  Its approximate solution
(\ref{eq:nscatter}) is obtained, again, by setting the derivative in
the left-hand side to zero.  This is true for
$d(a^3n_\mathrm{scatter})/dt \ll a^32B\omega n_\mathrm{scatter}$, that
is for $X\ll4B\xi\Xcr\approx4\times10^3\Xcr$, while we are interested
in much smaller $X$.  We should check of course, that the particle
density is not to small, to allow for stochastic resonance to work.
Again, for the typical occupation number (up to some numerical factor)
we get
\begin{equation}
  \label{eq:50}
  n_\bk
  \sim \frac{n_\mathrm{scatter}}{K^3}
  = \frac{2\,B}{5\,\pi^2\,\alpha_W^2}
  \approx 3.4
  \;.
\end{equation}
This is larger, than one, thus we may hope that exponential creation is
already a reasonable approximation for \eqref{eq:niterative}.  The
energy drain is obtain if we recall that the W bosons are
nonrelativistic, so each scattering provides the energy transfer of
$2\sqrt{\vev{m_W^2}}$, leading to (\ref{eq:49}).  The solution
of (\ref{eq:49}) is also saturated at late times, so (if the initial
time is small) we have
\begin{equation}
  \label{eq:51}
  \rho = \frac{96\,\xi\,\Xcr^{\frac{7}{2}}\,B^2\,\sqrt{X}\,\sqrt{
        \lambda}}{65\,\pi\,\sqrt{\alpha_W}}
  \approx 73 \, \l \frac{\lambda}{0.25}\r\,\sqrt{\Xcr^7 X}
  \;.
\end{equation}
One can see, that it reaches inflaton energy density
$\rho\sim\rho_\mathrm{inf}$ \eqref{eq:rhoinf} at
\begin{equation}
  \label{eq:52}
  X \sim \frac{16\,6^{\frac{2}{3}}\,\xi^{\frac{2}{3}}\,B^{\frac{4
          }{3}}}{65^{\frac{2}{3}}\,\pi^{\frac{2}{3}}\,
      \alpha_W^{\frac{1}{3}}\,\lambda^{\frac{1}{3}}} \Xcr
  \approx 110\Xcr
  \;.
\end{equation}
This is earlier, than the end of the non-resonant production region
\eqref{eq:42}.  Taking into account Z bosons makes this process
even more active.  This means, that after the moment \eqref{eq:42},
the parametric resonance starts, and due to higher concentration of
the gauge bosons, the energy is rapidly transferred into relativistic
SM fermions via gauge boson annihilation.  So, we expect that the
transfer of the energy to the relativistic modes via the gauge bosons
completes by approximately \eqref{eq:42}.

\section{Non-resonant Higgs production on the
  nonlinearities of the potential for small $\chi$}
\label{app:nr-h-production}

One needs to analyse the production of particles by
(\ref{eq:productioneq}), but with the mass
\begin{equation}
  m^2\l t\r =
  \left\{\begin{array}{l@{\;\;\;\text{for}\;}l}
      \omega^2
      & X\cos\left[\omega
        t\right]>\frac{\omega}{\sqrt{3\lambda}}
      \;,
      \\
      3\lambda X^2 \cos^2\left[\omega t\right] 
      & X\cos\left[\omega
        t\right]<\frac{\omega}{\sqrt{3\lambda}}
      \;.
    \end{array}
  \right.
\end{equation}
One way to find the generation by this source is to use the method
described in section \ref{app:semiclassic}.  The adiabatic
approximation holds while the mass does not change, and close to the
moments $t_j$ the problem can be solved after replacing the cosine
with the quadratic function.  Alternatively, if the number of
generated particles is small, a simpler perturbative approach can be
used.  We use the perturbative approach here.

Let us first neglect the expansion of the Universe during several
oscillations of the inflaton field.  In this case the number of
particle of the mass $\omega$ generated by the quadratic potential
\[
  \cL_{\mathrm{int}}=\frac{m^2\l t\r -\omega^2}{2}\l \delta \chi\r^2
  \;,
\]
is given by
\[
  n_\bk\l t\r =\left| \frac{1}{2k_0}\int_0^t dt \l m^2\l t\r
    -\omega^2\r \e^{2ik_0t}\right|^2
  \;,
\]
where $k_0^2=k^2+\omega^2$.  The integral is equal to (for the moment
of time around $t\sim t_l=\frac{2\pi}{\omega}l$, $l=1,2,\dots$; the
integral value changes while the inflaton field crosses zero, but the
exact form is not important)
\[
n_k\l t_l\r = \frac{1}{16}\frac{\sin^2\l 2l\frac{\pi
      k_0}{\omega}\r  }{\sin^2\l\frac{\pi k_0}{\omega}\r  }
L^2
\;,
\]
where
\begin{equation}
  L=\frac{3\lambda X^2}{2}\cdot \frac{\omega}{k_0} 
  \left\{
    \frac{1}{k_0+\omega} \sin\left[ 4\pi \l \frac{k_0}{\omega}
      +1\r \epsilon \right]
    - \frac{1}{k_0-\omega} 
    \sin\left[4\pi \l \frac{k_0}{\omega}-1\r \epsilon
    \right]
  \right\}
  \;,
\end{equation}
and the parameter $\epsilon$ is defined from the equation
\begin{equation}
  \label{Yepsilon}
  \sin\left[ 2\pi \epsilon\right]\equiv\frac{\omega}{\sqrt{3\lambda}X}
  \;.
\end{equation}
At large times, $t\gg\omega^{-1}$, using the equality
$\lim_{t\to\infty}\frac{\sin^2xt}{\pi x^2t}=\delta\l x\r $ one gets
\[
\frac{n_k\l t\r}{t}
\simeq\frac{\omega^2}{4\pi^2}\pi \sum_{l=1}^\infty L^2 \delta\l
      k_0-\omega l\r\;.
\]
Thus, integration over momenta gives a convergent sum
\begin{equation}
  \frac{n\l t \r}{ t} =
    \frac{9\lambda^2 X^4}{2^5\pi^3}\sum_{l=2}^\infty 
  \frac{1}{l^2}\sqrt{1-\frac{1}{l^2}} 
  \cdot 
  \left(
    \frac{1}{l+1} \sin\left[4\pi \l l+1\r \epsilon
    \right]
    - \frac{1}{l-1} 
    \sin\left[4\pi \l l-1\r \epsilon
    \right]
  \right)^2
  \;.
\end{equation}
The sum is saturated for $4\pi n \epsilon\simeq 1$. This implies the
typical energy of the produced particles,
\[
  E\sim\frac{\omega}{4\pi \epsilon}\sim\half \sqrt{3\lambda}X
  \;,
\]
which is larger, than $\omega$, so the particles are relativistic.

Using formulas from Appendix \ref{app:sums} we get the following
estimates for the production rate
\[
  \dot n\simeq \frac{n\l t \r}{ t} \simeq \frac{4\pi \omega^4}{15\pi^3} 
  \frac{\omega}{\sqrt{3\lambda} X}
  \;,
\]
and for the energy flux
\[
\dot\rho\simeq\frac{\rho\l t \r}{ t} \simeq \frac{\omega^5}{2\pi^3}\;. 
\]
Reintroducing the expansion of the Universe in the usual way by
changing $\dot n\to\frac{d(a^3n)}{a^3dt}$,
$\dot\rho\to\frac{d(a^4\rho)}{a^4dt}$,
we have the number and energy densities at late time
\begin{gather}
  n = \frac{4\pi\omega^4}{15\pi^3} \frac{\omega}{4}
  \frac{t}{\sqrt{3\lambda}X(t)}
  \;,
  \\
  \rho = \frac{3}{11}\frac{\omega^5}{2\pi^3}t
  \;.
\end{gather}

\section{Tunnelling through a $-|x|$ barrier}
\label{app:modulebarrier}

The solution of eq.~\eqref{eq:17} is given by the Airy
functions for negative and positive times:
\begin{align}
  \label{eq:32}
  \varPhi_\bk(\tau<0) &= A_-\Ai(\tau-\kappa^2)+B_-\Bi(\tau-\kappa^2)
  \;,\\
  \varPhi_\bk(\tau>0) &= A_+\Ai(-\tau-\kappa^2)+B_+\Bi(-\tau-\kappa^2)
  \;.
\end{align}
The coefficients should be determined by the matching conditions at
$\tau=0$
\begin{align}
  \label{eq:34}
  \varPhi_\bk(0-) &= \varPhi_\bk(0+)
  \;, &
  \varPhi'_\bk(0-) &= \varPhi'_\bk(0+)
  \;.
\end{align}
It is comfortable, however, firstly to match the coefficients $A_\pm$,
$B_\pm$ with $\alpha=\alpha_\bk^j\e^{-i\theta_\bk^j}$,
$\beta=\beta_\bk^j\e^{+i\theta_\bk^j}$,
$\alpha'=\alpha_\bk^{j+1}\e^{-i\theta_\bk^j}$, and
$\beta'=\beta_\bk^{j+1}\e^{i\theta_\bk^j}$ from the expansion
\eqref{eq:13}.  The asymptotic expansions of the Airy functions are
\begin{align}
  \label{eq:33}
  \Ai(-x) & = \frac{1}{\sqrt{\pi}x^{1/4}}\sin\left(
    \frac{2}{3}x^{3/2}+\frac{\pi}{4}
  \right)
  \;,\\
  \Bi(-x) & = \frac{1}{\sqrt{\pi}x^{1/4}}\cos\left(
    \frac{2}{3}x^{3/2}+\frac{\pi}{4}
  \right)
  \;.
\end{align}
Then, the solution matched with \eqref{eq:13} at large $\tau$ (for
matching one should use in \eqref{eq:13} only linear part of the mass,
$m_W(t)\simeq \const \cdot |t-t_j|$)
\begin{align}
  \varPhi_\bk(\tau<0) = &\,
  \sqrt{\frac{\pi}{2}}\left[
    \alpha\e^{-i(\frac{2}{3}\kappa^3+\frac{\pi}{4})}
    +\beta\e^{i(\frac{2}{3}\kappa^3+\frac{\pi}{4})}
  \right]\Bi(\tau-\kappa^2) +
\\\notag  &
+i\sqrt{\frac{\pi}{2}}\left[
    \alpha\e^{-i(\frac{2}{3}\kappa^3+\frac{\pi}{4})}
    -\beta\e^{i(\frac{2}{3}\kappa^3+\frac{\pi}{4})}
  \right]\Ai(\tau-\kappa^2)
  \;,\\
  \varPhi_\bk(\tau>0) =&\,
  \sqrt{\frac{\pi}{2}}\left[
    \alpha'\e^{i(\frac{2}{3}\kappa^3+\frac{\pi}{4})}
    +\beta'\e^{-i(\frac{2}{3}\kappa^3+\frac{\pi}{4})}
  \right]\Bi(-\tau-\kappa^2)+
\\\notag  &
+i\sqrt{\frac{\pi}{2}}\left[
    -\alpha'\e^{i(\frac{2}{3}\kappa^3+\frac{\pi}{4})}
    +\beta'\e^{-i(\frac{2}{3}\kappa^3+\frac{\pi}{4})}
  \right]\Ai(-\tau-\kappa^2)
  \;.
\end{align}
Using the linear relations between $\alpha$, $\beta$, $\alpha'$,
$\beta'$ from condition at zero \eqref{eq:34}, and the definition of
$R_\bk$, $D_\bk$ from \eqref{eq:14}, which is
\begin{align}
  \label{eq:36}
  \alpha' &= \alpha\frac{1}{D_\bk}+\beta\frac{R_\bk^*}{D_\bk^*}
  \;,\\
  \beta' &= \alpha\frac{R_\bk}{D_\bk}+\beta\frac{1}{D_\bk^*}
  \;,
\end{align}
one gets the expressions \eqref{eq:18}, \eqref{eq:18'}.

\section{Useful sums}
\label{app:sums}

In Appendix \ref{app:nr-h-production} we obtain the following sums 
\begin{align}
S_1&=\sum_{l=2}^\infty 
\frac{1}{l^2}\sqrt{1-\frac{1}{l^2}} 
\cdot 
\l
\frac{1}{l+1} \sin\left[4\pi \l l+1\r \epsilon
\right] - \frac{1}{l-1} 
\sin\left[4\pi \l l-1\r \epsilon
\right]
\r^2\;,\\
S_2&=
\sum_{l=2}^\infty 
\frac{1}{l}\sqrt{1-\frac{1}{l^2}} 
\cdot 
\l
\frac{1}{l+1} \sin\left[4\pi \l l+1\r \epsilon
\right] - \frac{1}{l-1} 
\sin\left[4\pi \l l-1\r \epsilon
\right]
\r^2\;,
\end{align}
where $\epsilon$ is a small dimensionless parameter, $\epsilon\ll
1$. These sums are saturated at $l\sim1/4\pi\epsilon$, hence to get
the leading order results in $\epsilon$ one can replace these sums
with corresponding integrals by introducing a new variable $u$ as
$u=4\pi l \epsilon$, $du=4\pi\epsilon$. Then to the leading order in
$\epsilon$ one arrives at
\begin{align}
S_1&= \int_0^\infty \epsilon^5  
\frac{du}{u^6}
\l
2 u \cos u -2 \sin u \r^2 
= \frac{4\pi}{15}\l 4\pi \epsilon\r^5\;,\\
S_2&= \int_0^\infty \epsilon^4 
\frac{du}{u^5}
\l
2 u \cos u -2 \sin u \r^2 
= \l 4\pi \epsilon\r^4\;.
\end{align}

\section{Nonresonant particle production with $h$ and $|h|$ sources}
\label{app:|h|source}

Here we compare production of fermions by $h$ and $|h|$ sources and
conclude, that the corresponding production rates are the same, though
the spectra differ.

As far as the number of the created particles is small, $n_\bk\ll1$,
we can use the perturbation theory.  Then, the perturbation
\[
  \hat H_{int} \equiv \int d^3{\mathbf{x}} m\l t\r \bar\Psi \Psi
\]
leads to the number density at the moment t
\[
  n_k\l t \r = \int_0^t m\l t'\r \e^{2ik_0t'} dt'
               \int _0^t m\l t''\r \e^{-2ik_0t''} d t''
  \;.
\]
and total particle number
\[
  n(t)=\int \frac{d^3\bk}{(2\pi)^3}n_k(t)
  \;.
\]

Here we calculate the particle density
side by side for two different sources
\begin{align}
  \label{eq:22}
  m(t) &= m\sin(\omega t) \;, \\
  m(t) &= |m\sin(\omega t)| \;,
 \label{eq:22b}
\end{align}
and massless fields $\bar \Psi$, $\Psi$, so $k_0=k$.  
One can check that for $n=0,1,\dots$ 
\begin{align}
  \int_{\frac{2\pi n}{\omega}}^{\frac{2\pi\l n+\half\r}{\omega}} 
  dt\sin\l \omega
  t\r\e^{2ikt}
  &=
  -\frac{2\omega}{4k^2-\omega^2}\cdot\e^{i\frac{\pi k}{\omega}} 
  \cos\l \frac{k\pi}{\omega}\r\cdot \e^{i\frac{4\pi kn}{\omega}}\;, \\
  \int_{\frac{2\pi\l n+\half\r}{\omega}}^{\frac{2\pi\l n+1\r}{\omega}} 
  dt\sin\l \omega
  t\r\e^{2ikt}
  &=
  -\frac{2\omega}{4k^2-\omega^2}\cdot\e^{3i\frac{\pi k}{\omega}} 
  \cos\l \frac{k\pi}{\omega}\r\cdot \e^{i\frac{4\pi kn}{\omega}}\;.
\end{align}
Hence for the full $\l n+1\r $th period  
\begin{align}
  \int_{\frac{2\pi n}{\omega}}^{\frac{2\pi\l n+1\r}{\omega}} 
  dt\sin\l \omega
  t\r\e^{2ikt}
  &=
  \frac{4i\omega}{4k^2-\omega^2}\cdot\e^{i\frac{2\pi k}{\omega}} 
  \cos\l \frac{k\pi}{\omega}\r\cdot \sin\l \frac{k\pi}{\omega}\r
  \e^{i\frac{4\pi kn}{\omega}} \;,\\
  \int_{\frac{2\pi n}{\omega}}^{\frac{2\pi\l n+1\r}{\omega}} 
  dt\left|\sin\l \omega
    t\r\right|\e^{2ikt}
  &=
  \frac{-4\omega}{4k^2-\omega^2}\cdot
  \e^{i\frac{2\pi k}{\omega}} 
  \cos^2\l \frac{k\pi}{\omega}\r\cdot 
  \e^{i\frac{4\pi kn}{\omega}}\;,
\end{align}
and summing over $N+1$ periods one arrives at 
\begin{align}
  \int_0^\frac{2\pi \l N+1\r}{\omega}dt\;\sin(\omega t)\, \e^{2ik_0t}=
  \frac{4i\omega}{4k^2-\omega^2}
  \cdot\e^{i\frac{2\pi k\l N+1\r}{\omega}} 
  \cdot\cos\l \frac{k\pi}{\omega}\r\cdot \sin\l \frac{k\pi}{\omega}\r
  \cdot\frac{\sin\l \frac{2\pi k\l N+1\r}{\omega}\r}{\sin\l 
    \frac{2\pi k}{\omega}\r }\;, \\
  \int_0^\frac{2\pi \l N+1\r}{\omega}dt\;\left|\sin(\omega t)\right| 
  \e^{2ik_0t}= 
  \frac{4i\omega}{4k^2-\omega^2}
  \cdot\e^{i\frac{2\pi k\l N+1\r}{\omega}} 
  \cdot\cos^2\l \frac{k\pi}{\omega}\r\cdot 
  \frac{\sin\l \frac{2\pi k\l N+1\r}{\omega}\r}{\sin\l 
    \frac{2\pi k}{\omega}\r }\;.
\end{align}
Hence the number of produced particles for $N+1$ periods of oscillations
is 
\begin{align}
  n\l t=\frac{2\pi\l N+1\r}{\omega}  \r & 
  = \frac{m^2}{2\pi^2}\int dk
  \frac{16\omega^2k^2}{\l 4k^2-\omega^2\r^2} 
  \cos^2\l \frac{k\pi}{\omega}\r\cdot \sin^2\l \frac{k\pi}{\omega}\r
  \frac{\sin^2\l \frac{2\pi k\l N+1\r}{\omega}\r}{\sin^2\l 
    \frac{2\pi k}{\omega}\r }\;, \\
  n\l t=\frac{2\pi\l N+1\r}{\omega}  \r & 
  = \frac{m^2}{2\pi^2}\int dk
  \frac{16\omega^2k^2}{\l 4k^2-\omega^2\r^2} 
  \cos^4\l \frac{k\pi}{\omega}\r\cdot 
  \frac{\sin^2\l \frac{2\pi k\l N+1\r}{\omega}\r}{\sin^2\l 
    \frac{2\pi k}{\omega}\r }\;.
\end{align}

Assuming that a tiny amount of particles is produced per each period
one can turn to continuous variable $T$ in these expressions.  To
obtain the particle production rate we are interested in linear in $T$
contribution. It comes from poles in the integrands. Having this in
mind and making use of the relation
\begin{equation}
\label{QuiteUsefulRelation}
\lim_{t\to\infty}\frac{\sin^2\alpha t}{\pi t \alpha^2}=\delta\l
\alpha \r\;,
\end{equation}
one proceeds with calculations. For the source \eqref{eq:22}
one makes use of the identity
\[
\sin^2\l \frac{2\pi k\l N+1\r}{\omega}\r=
\sin^2\l \l k-\frac{\omega}{2}\r t\r\;,
\]
while for the source \eqref{eq:22b}: 
\[
\sin^2\l \frac{2\pi k\l N+1\r}{\omega}\r=
\sin^2\l k T\r\;.
\]
Then for the first source one obtains
\[
n\l t \r =\frac{m^2}{2\pi^2}\int \frac{dk \; k^2}{\l
  k+\frac{\omega}{2}\r^2}\frac{\omega^2}{4} 
\frac{\sin^2\l Tk\r}{\l k-\frac{\omega}{2}\r^2}
\]
and with help of \eqref{QuiteUsefulRelation} 
\[
n\l t \r =t\frac{m^2}{2\pi^2}\int \frac{dk \; \pi k^2}{\l
  k+\frac{\omega}{2}\r^2}\frac{\omega^2}{4} 
\delta\l k-\frac{\omega}{2}\r\;.
\]
This is the monochromatic spectrum. The number of produced particles
is 
\[
n\l t \r = t\cdot \frac{m^2\omega^2}{32\pi}\;.
\]

For the second source \eqref{eq:22b} one has 
\[
n\l t \r =\frac{m^2}{2\pi^2}\int \frac{dk \; k^2}{\l
  k+\frac{\omega}{2}\r^2}\frac{\omega^2}{4} 
\frac{\cos^2\l \frac{k\pi}{\omega}\r}{\l k-\frac{\omega}{2}\r^2}
\frac{\sin^2\l Tk\r}{\sin^2\l \frac{k\pi}{\omega}\r}\;.
\]
Here the double-pole at $k=\omega/2$ is cancelled by the double-zero
from cosine squared. But there are a lot of poles due to sine in the
denominator.  In this case the useful variant of
\eqref{QuiteUsefulRelation} is 
\[
\lim_{t\to\infty}\frac{\sin^2\l tk\r}{\pi t \sin^2\l
  \frac{k\pi}{\omega}\r} = \frac{\omega^2}{\pi^2}\delta\l k\r\;.
\]
It gives for the spectra
\[
n\l t \r =t\frac{m^2}{2\pi^2}\cdot\frac{\omega^2}{\pi}
  \int \frac{dk \; k^2}{\l
  k+\frac{\omega}{2}\r^2}\frac{\omega^2}{4} 
\frac{\sum_n\delta\l k-\omega n\r}{\l k-\frac{\omega}{2}\r^2} 
\]
Finally, integrating over momenta and summing up the series
\[
\sum_{n=0}^{n=\infty}\frac{n^2}{\l n^2-\frac{1}{4}\r^2}=\frac{\pi^2}{4}
\]
one gets 
\[
n\l t \r = t\cdot \frac{m^2\omega^2}{32\pi}\;,
\]
the same answer as for the first source \eqref{eq:22}.

\bibliographystyle{JCAP-hyper}
\bibliography{all}

\end{document}